\documentclass[reprint, amsmath,amssymb, aps ]{revtex4-2}

\usepackage{graphicx}% Include figure files
\usepackage{dcolumn}% Align table columns on decimal point
\usepackage{bm}% bold math
\usepackage{xcolor}
\usepackage{physics}
\usepackage{mathtools}
\usepackage[normalem]{ulem}

\begin{document}
	
	\title{Role of Multiple Scattering in Single Particle Perturbations in Absorbing Random Media}
	
	\author{Joel Berk}
	\author{Matthew R. Foreman}%
	\email{matthew.foreman@imperial.ac.uk}
	\affiliation{Blackett Laboratory, Imperial College London, Prince Consort Road, London, SW7 2BW, United Kingdom}
	
	\date{\today}
	
	\begin{abstract}
		Speckle patterns produced by disordered scattering systems exhibit a sensitivity to addition of individual particles which can be used for sensing applications. Using a coupled dipole model we investigate how multiple scattering can enhance field perturbations arising in such random scattering based sensors. Three distinct families of multiple scattering paths are shown to contribute and the corresponding complex enhancement factors derived. Probability distributions of individual enhancement factors over the complex plane are characterised numerically within the context of surface plasmon polariton scattering in which absorption is shown to play an important role.   We show that enhancements become more strongly dependent on individual scatterer properties when absorption losses are larger, however, amplitude enhancements $\sim 10^2$, comparable to low loss surface plasmons, are achievable through sensor optimisation. Approximate analytic expressions for the complex mean enhancements are also found, which agree well with simulations when loop contributions are negligible.
	\end{abstract}
	
	
	\maketitle

	\section{Introduction} 
	Use of optical scattering for detection and measurement is a powerful and widespread approach underpinning techniques such as interferometric scattering microscopy (iSCAT) \cite{Taylor2019InterferometricScattering}, dynamic light scattering \cite{Stetefeld2016DynamicSciences} and diffusing wave spectroscopy \cite{DWS88}. This phenomenon has seen extensive application in the biological sciences and environmental monitoring, in turn driving development of scattering based sensors. As sensitivity gains have been made, so sensing has moved from monitoring of bulk properties to detection of individual nanometer-sized analyte particles, such as virions and proteins \cite{Ye2019}. Such small dimensions however mean particles only scatter weakly, hence presenting a major challenge. To mitigate this issue strongly confined fields, which can enhance light-matter interactions, have been employed, for example, using high Q optical resonators \cite{Shao2014}, photonic crystals \cite{Li2021} and nanoapertures \cite{Xue2020}. Plasmonic systems, supporting localized or propagating surface plasmon-polaritons (SPPs), are also particularly attractive for sensing, since in addition to confining optical fields they can be easily implemented on chip scale devices, are biocompatible, allow operation in aqueous/microfluidic environments and can exploit the existing wealth of functionalisation protocols required to maintain specificity \cite{Homola2003PresentBiosensors,Baldrich2008}. Accordingly, SPPs have found applications in numerous sensing and particle tracking setups  \cite{JeffreyN.Anker2008BiosensingNanosensors,Zhang2020PlasmonicKinetics,Zijlstra2012OpticalNanorod,Raschke2003,Taylor2017,Xue2019}. 
	
	Interferometric plasmonic systems provide one route to yet further sensitivity gains \cite{Wen2015,Feng2012a,Bian2013,Zeng2015,Yang2018InterferometricExosomes} by leveraging coherent detection. Alternatively, nanostructured sensors, such as nanoparticle arrays, aperiodic gratings and randomly disordered substrates \cite{Enoch2004, DalNegro2012,Lee2010,LeMoal2009, Szunerits2008} have also shown significant promise. 
	Random scattering, in particular, affords numerous opportunities in sensing by virtue of the diverse range of phenomena that can occur. For example, depending on the degree of multiple scattering (as parametrized by the scattering mean free path), scattering can give rise to long and short range correlations, weak and strong (Anderson) localisation and fluctuations in the local density of states. A substantial amount of work has been dedicated to study such phenomena in both the optical \cite{Berkovits1994,Maystre1994,Boguslawski2017,Segev2013,Shapiro1999NewMedia,Skipetrov2000NonuniversalScattering} and plasmonic domains \cite{VanBeijnum2012b,Arnold1996,Bozhevolnyi1996bbb,Caze2012,Carminati2015, Foreman2019a, Bozhevolnyi2007} over the years. Indeed, exploitation of random scattering has a rich track record in optics. For example, correlations present in the speckle patterns have been used for refractive index sensing, spectrometry and  imaging \cite{Tran2020UtilizingSensing,MaumitaChakrabarti2015, Katz2014, Lee2016}. Speckle patterns generated by disordered multiple scattering environments have also been shown to depend on the properties of individual scatterers \cite{Nieuwenhuizen1993,Berkovits1991}, such as their position, effective charge or orientation \cite{Berkovits1990, Lancaster1998,Vynck2014}, whilst also providing enhanced sensitivity as compared to single scattering environments \cite{Berk2021}. Approaches to extract the position of a single scatterer accounting for multiple scattering effects have thus been developed, for example based on diffusive models of light propagation \cite{denOuter:93} or extension of single scattering holography localisation techniques \cite{MultScattHolographLocalization,Suski2020FastMethod}. Recent advances in machine learning moreover present further opportunities to extract information from randomly scattered light, since such approaches do not require a detailed physical model and are hence applicable across a broad range of scattering regimes \cite{Kamilov2015LearningTomography,Sun2018EfficientLearning,Moon2019}.
	
	 In plasmonics, random scattering has also seen employ, for example, in light harvesting, super-resolution imaging and sensing \cite{Nishijima:12, Kim2010,Szunerits2008,Perumal2014,Frolov2013}. Absorption associated with Ohmic losses in metals is, however, intrinsic to plasmonic systems \cite{Khurgin2015HowMetamaterials}. For resonance tracking based sensors, absorption broadens the resonance lineshape and thus limits sensitivity. Statistical properties of speckle patterns in absorbing multiple scattering environments can however also be affected, for example, absorption can give rise to non-Rayleigh intensity statistics, as well as generate non-universal and reduced correlations \cite{Sangu1999EffectMedia,Genack1993IntensityMedia,Pnini1991,Foreman2019a}. In this article, we address the open question as to how absorption affects sensitivity in random scattering based particle sensing. Particular emphasis is placed on surface plasmon based setups due to their prevalence and capabilities in this domain.	To address this question, in Sections~\ref{sec:theory1} and \ref{sec:theory2} we first derive three enhancement factors, arising from three distinct categories of multiple scattering paths, which describe the effect of multiple scattering on the electric field perturbation caused by the presence of an additional analyte particle. We recently studied the \emph{amplitude} of these enhancement factors in the context of multiple scattering of SPPs  \cite{Berk2021} by randomly distributed scatterers on a metal surface, however here we study the full probability distribution of the enhancements on the complex plane, including phase effects. Approximate analytic results for the mean enhancement factors are derived in Section~\ref{sec:theory3}, before numerical results are given in Section~\ref{sec:numerical}. The dependence of the achievable enhancements and the associated probability distributions on absorption loss is detailed in Section~\ref{sec:numerical2}. In particular, through consideration of the role of scattering phase, propagation phase and absorption we identify a non-trivial dependence of the mean enhancement on tunable properties of the scattering configuration. This dependence is explored as a route to sensor optimisation in Section~\ref{sec:numerical3}. As such the insights gained in this work allow us not only to understand the interplay of absorption and multiple scattering upon addition of an analyte particle, but also to guide future development of optimal random SPP sensors.

	\section{Theory}\label{sec:enhancements}
	\subsection{Coupled Dipole Model}\label{sec:theory1}
	The disordered scattering system we shall study is a collection of $N$ coupled point dipole scatterers \cite{Novotny1997InterferencePlasmons,Chaumet2005EfficientMethod,Sndergaard2003VectorialInteractions}, situated in an environment with background dielectric function $\varepsilon(\bm{r})$. A Green's tensor $G(\bm{r},\bm{r}')$ can be defined for this system as the solution to Maxwell's wave equation
	\begin{equation}\label{eq:greens_tensor_def}
		\curl\curl G(\bm{r},\bm{r}')-\varepsilon(\bm{r})k_0^2G(\bm{r},\bm{r}')=I\delta(\bm{r}-\bm{r}'),
	\end{equation}
	where $k_0=\omega/c=2\pi/\lambda_0$, $\omega$ is the angular frequency, $c$ is the speed of light, $\lambda_0$ is the wavelength in vacuum and $I$ is the $3\times3$ identity matrix. When the point scatterers are illuminated with a monochromatic incident electric field $\bm{E}_0(\bm{r})$, the total electric field $\bm{E}(\bm{r})$ at position $\bm{r}$ is 
	\begin{equation}\label{eq:coup_dip}
		\bm{E}(\bm{r})=\bm{E}_0(\bm{r})+\frac{k_0^2}{\varepsilon_0}\sum_{j=1}^N G(\bm{r},\bm{r}_j)  \bm{p}_j,
	\end{equation}
	where $\bm{r}_j$, $\alpha_j$ and {$\bm{p}_j=\alpha_j\bm{E}_{\text{exc}}(\bm{r}_j)$} are the position,  dressed polarizability and dipole moment of the $j$th scatterer respectively, {and $\bm{E}_\text{exc}(\bm{r}_j)=\bm{E}_0(\bm{r}_j)+\sum_{i\neq j}G(\bm{r}_j,\bm{r}_i)  \bm{p}_i$ is the exciting field incident on the $j$th dipole, consisting of the incident field and the field from all other dipoles \cite{Novotny1997InterferencePlasmons,LakhtakiaCDA}}. Notably, $\alpha_j$ includes the effect of self-interactions (e.g. due to reflections from the background medium). From Eq.~\eqref{eq:coup_dip} we can construct the set of linear equations 
	\begin{equation}\label{eq:Mp_p0}
		\sum_{j=1}^NM_{ij}\bm{p}_j=\bm{p}_{0,i},
	\end{equation}
	for $i = 1,2,\ldots N$, where $\bm{p}_{0,i}=\alpha_i\bm{E}_0(\bm{r}_i)$ is the dipole moment induced by the incident field in the $i$th scatterer, the matrix elements $M_{ij}$ are defined by
	\begin{equation}\label{eq:M_def}
		M_{ij}=\begin{cases}
			I\quad &i=j\\
			-\frac{k_0^2}{\varepsilon_0}\alpha_iG_{ij} \quad &i\neq j,
		\end{cases}
	\end{equation}
	for $i,j = 1,2,\ldots N$, and $G_{ij}=G(\bm{r}_i,\bm{r}_j)$. Once Eq.~\eqref{eq:Mp_p0} is solved for the $N$ dipole moments, the field at any point can be calculated using Eq.~\eqref{eq:coup_dip}.
	Throughout this analysis, we consider scattering of a vector field with corresponding Green's tensor, such that $M_{ij}$ are the tensor elements of an $N\times N$ matrix of tensors (or equivalently they are the $3\times3$ blocks making up a $3N\times 3N$ matrix), which we denote ${M}$. Our analysis, however, is equally valid for scattering of a scalar field, if $G$, $\alpha$, $\bm{E}$ and $\bm{p}_i$ are replaced with scalar equivalents, in which case $M_{ij}$ are the scalar elements of an $N\times N$ matrix. For random positions, $\bm{r}_i$, the matrix ${M}$ is a Euclidean random matrix, the statistics of which have been studied, for example, in the context of optical scattering and vibrational modes of glasses \cite{Goetschy2011Non-HermitianTheory,Goetschy2013EuclideanPhysics,Martin-Mayor2001TheSystems,Mezard1999SpectraMatrices}. Within the single scattering regime, the off-diagonal terms describing coupling between the dipoles are negligible such that $M_{ij}\approx I\delta_{ij}$ and $\bm{p}_i\approx\bm{p}_{0,i}$.

	\subsection{Adding a Scatterer}\label{sec:theory2}
	We now consider perturbing the scattering configuration by introducing an additional point scatterer with polarizability $\alpha_{N+1}$ at position $\bm{r}_{N+1}$. The perturbed system can be described similarly to above yielding the set of coupled dipole equations $\sum_{j=1}^{N+1}M_{ij}'\bm{p}_{j}'=\bm{p}_{0,i}$ ($i=1,2,\ldots N+1$) in terms of the modified dipole moments $\bm{p}_j'$. We note that the matrix elements $M_{ij}'$ for the perturbed system are again given by Eq.~\eqref{eq:M_def} albeit the indices $i$ and $j$ run from $1$ to $N+1$ (hence $M_{ij}'=M_{ij}$, for $i,j\leq N$). 
	The new set of $N+1$ dipole moments results in the perturbed field $\bm{E}'$ (cf. Eq.~\eqref{eq:coup_dip})
	\begin{equation}\label{eq:e2}
		\bm{E}'(\bm{r})=\bm{E}_0(\bm{r})+\frac{k_0^2}{\varepsilon_0}\sum_{j=1}^{N+1} G(\bm{r},\bm{r}_j)  \bm{p}'_j. 
	\end{equation}
	Accordingly, the perturbation to the field $\delta\bm{E}=\bm{E}'-\bm{E}$ caused by the addition of the scatterer is hence
	\begin{equation}\label{eq:delta_E}
		\delta\bm{E}(\bm{r})=\frac{k_0^2}{\varepsilon_0}{G}(\bm{r},\bm{r}_{N+1}) \bm{p}_{N+1}+\frac{k_0^2}{\varepsilon_0}\sum_{j=1}^N {G}(\bm{r},\bm{r}_j)  \delta \bm{p}_j,
	\end{equation}
	where $\delta\bm{p}_j=\bm{p}'_j-\bm{p}_j$ is the perturbation to the $j$th dipole moment and, since there is no $(N+1)$th scatterer in the unperturbed system, we have dropped the prime from $\bm{p}_{N+1}$. The first term of Eq.~\eqref{eq:delta_E} corresponds to the field scattered by the added dipole $\bm{p}_{N+1}$, whereas the second term arises because  multiple scattering introduces dipole coupling whereby the presence of the additional scatterer modifies the $N$ initial dipole moments. In the single scattering regime, the coupling between dipoles is negligible so that $\delta\bm{p}_i=\bm{0}$ and the second term vanishes. Similarly, $\bm{p}_{N+1}=\bm{p}_{0,N+1}$ such that the single scattering perturbation $\delta\bm{E}_{ss}(\bm{r})$ reduces to
	\begin{equation}\label{eq:delta_E_ss}
		\delta\bm{E}_{ss}(\bm{r})=\frac{k_0^2}{\varepsilon_0}{G}(\bm{r},\bm{r}_{N+1}) \bm{p}_{0,N+1}.
	\end{equation}
	
	The coupled dipole equations for the perturbed $N+1$ scatterer system can be expressed in the form
	\begin{align}
		\sum_{j=1}^NM_{ij}(\bm{p}_j+\delta\bm{p}_j)-\frac{k_0^2}{\varepsilon_0}\alpha_{i}G_{i,N+1}\bm{p}_{N+1}&=\bm{p}_{0,i}, \label{eq:dp_eq}\\
		\bm{p}_{N+1}-\frac{k_0^2}{\varepsilon_0}\sum_{j=1}^N\alpha_{N+1} {G}_{N+1,j} (\bm{p}_j+\delta \bm{p}_j)&=\bm{p}_{0,N+1}. \label{eq:p_tot_eq}
	\end{align}
	Using Eq.~\eqref{eq:Mp_p0}, Eq.~\eqref{eq:dp_eq} can in turn be rearranged to yield
	\begin{equation}\label{eq:delta_p_solution}
		\delta \bm{p}_i=\sum_{j=1}^N{M}_{ij}^{-1} \frac{k_0^2}{\varepsilon_0}\alpha_j {G}_{j,N+1} \bm{p}_{N+1},
	\end{equation}
	where ${M}_{ij}^{-1}$ is here used to denote the the $(i,j)$th $3\times3$ block (corresponding to rows $3i-2$ to $3i$ and columns $3j-2$ to $3j$) of the inverse of the entire $3N\times3N$ matrix ${M}$, as opposed to $(M_{ij})^{-1}$, the inverse of the $3\times3$ sub-matrix $M_{ij}$ (similarly, in the scalar case, it corresponds to the $(i,j)$th element of the inverse of the $N\times N$ matrix ${M}$). Substituting Eq.~\eqref{eq:delta_p_solution} into Eq.~\eqref{eq:delta_E} then gives
	\begin{equation}
		\delta\bm{E}(\bm{r})=\frac{k_0^2}{\varepsilon_0}{G}(\bm{r},\bm{r}_{N+1})\gamma_1(\bm{r}) \bm{p}_{N+1}.
	\end{equation}
	where we have also defined the enhancement factor
	\begin{equation}
		\gamma_1(\bm{r})=I+\frac{k_0^2}{\varepsilon_0}G(\bm{r},\bm{r}_{N+1})^{-1}\sum_{i,j=1}^N {G}(\bm{r},\bm{r}_i){M}_{ij}^{-1} \alpha_j {G}_{j,N+1}.\label{eq:gamma_1}
	\end{equation}
	Expressing $\delta\bm{E}$ as such allows comparison with the single scattering result in Eq.~\eqref{eq:delta_E_ss}. Specifically, it is evident that the perturbation to the dipole moments of the $N$ initial scatterers from introduction of an additional scatterer is described by the  factor $\gamma_1$. Equivalently, dipole coupling through multiple scattering acts to modify the effective dipole moment of the additional scatterer such that $\bm{p}_{N+1}\to\gamma_1(\bm{r}) \bm{p}_{N+1}$. The tensor nature of $\gamma_1$ reflects the fact that the polarization of the field perturbation can be modified by multiple scattering. Similarly, $\gamma_1$ is a complex quantity, implying multiple scattering can affect both the phase and amplitude of $\delta\bm{E}$.

	In addition to the dipole coupling captured in $\gamma_1$, there remain further multiple scattering effects which cause $\bm{p}_{N+1}\neq\bm{p}_{0,N+1}$. Specifically, the local field experienced by the additional scatterer is not solely dictated by the incident field $\bm{E}_0$, but also contains a contribution from scattering of the illumination field by the $N$ initial scatterers. To demonstrate this, we substitute Eq.~\eqref{eq:delta_p_solution} and Eq.~\eqref{eq:Mp_p0} into Eq.~\eqref{eq:p_tot_eq}, which results in
	\begin{align}
		&\bm{p}_{N+1}=\bm{p}_{0,N+1}+\frac{k_0^2}{\varepsilon_0}\sum_{i,j=1}^N\alpha_{N+1}  {G}_{N+1,i}  {M}_{ij}^{-1} \bm{p}_{0,j}\nonumber\\
		&\,+\left(\frac{k_0^2}{\varepsilon_0}\right)^2\sum_{i,j=1}^N\alpha_{N+1}  {G}_{N+1,i}  {M}_{ij}^{-1} \alpha_j  {G}_{j,N+1} \bm{p}_{N+1}.
	\end{align}
	Rearranging for $\bm{p}_{N+1}$ yields
	\begin{align}\label{eq:pN1_final}
		\bm{p}_{N+1}=&\Bigg[ {I}-\!\left(\frac{k_0^2}{\varepsilon_0}\right)^2\!\!\sum_{i,j=1}^N\!\alpha_{N+1}  {G}_{N+1,i}  {M}_{ij}^{-1} \alpha_j  {G}_{j,N+1}\Bigg]^{-1}\nonumber\\
		&\times\Bigg[\bm{p}_{0,N+1}+\frac{k_0^2}{\varepsilon_0}\sum_{i,j=1}^N\alpha_{N+1}  {G}_{N+1,i}  {M}_{ij}^{-1} \bm{p}_{0,j}\Bigg].
	\end{align}
	Defining two further enhancement factors allows $\delta\bm{E}$ to be expressed as
	\begin{equation}\label{eq:delta_E_ms}
		\delta\bm{E}(\bm{r})=\frac{k_0^2}{\varepsilon_0}{G}(\bm{r},\bm{r}_{N+1}) \gamma_1\gamma_2\gamma_3\bm{p}_{0,N+1},
	\end{equation}
	where
	\begin{align} 
		\gamma_2&=\Bigg[I-\!\left(\frac{k_0^2}{\varepsilon_0}\right)^2\!\!\sum_{i,j=1}^N\!\alpha_{N+1}  {G}_{N+1,i}  {M}_{ij}^{-1} \alpha_j  {G}_{j,N+1}\Bigg]^{-1}\label{eq:gamma_2}\\
		\gamma_3&=I+\frac{k_0^2}{\varepsilon_0}\sum_{i,j=1}^N\alpha_{N+1}  {G}_{N+1,i}  {M}_{ij}^{-1} \frac{\bm{p}_{0,i} \bm{p}_{0,N+1}^\dag}{ \abs{\bm{p}_{0,N+1}}^2}\label{eq:gamma_3}.
	\end{align}
	Expressed in this way, it can be seen that the effect of multiple scattering is equivalent to changing the dipole moment from $\bm{p}_{0,N+1}$ to $\gamma_1\gamma_2\gamma_3\bm{p}_{0,N+1}$. In general, as with $\gamma_1$, the enhancement factors $\gamma_{2}$ and $\gamma_3$ are complex matrices, meaning multiple scattering can change the phase, amplitude and polarization of $\delta\bm{E}$. 
	
\begin{figure}[t]
		\centering
		\includegraphics[width=\columnwidth]{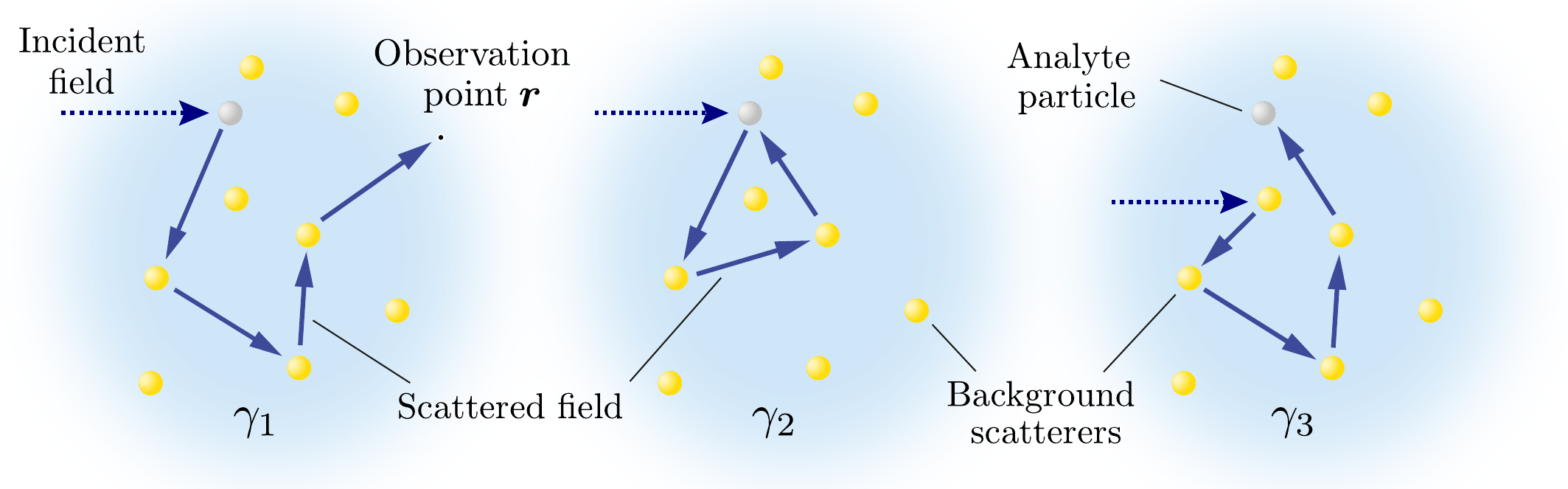}
		\caption{Example multiple scattering paths for each enhancement factor: ($\gamma_1$, left) rescattering  between scattering from the additional particle and propagation to the observation point, ($\gamma_2$, center) loop trajectories starting and ending on the additional scatterer and ($\gamma_3$, right) multiple scattering of the illumination field onto the analyte particle.}
		\label{fig:scattering_paths}
	\end{figure}

	Each enhancement factor can be associated with a class of multiple scattering paths involving the additional scatterer as shown in Fig.~\ref{fig:scattering_paths}. Firstly, the effect of rescattering of the field as it propagates to the observation point $\bm{r}$ after being scattered by the additional scatterer is accounted for by $\gamma_1$. The factor of $\alpha_jG_{j,N+1}$ freely propagates the scattered field from $\bm{r}_{N+1}$ to a scattering event at the $j$th scatterer, while $M_{ij}^{-1}$ propagates the field from the $j$th scatterer to the $i$th scatterer via all possible scattering paths involving the initial $N$ scatterers. Free propagation from the $i$th scatterer to the observation point $\bm{r}$ is then described by $G(\bm{r},\bm{r}_i)$. Secondly, the $\gamma_2$ factor describes the effect of loop scattering paths in which waves, after being scattered by the additional dipole, return (possibly multiple times) to the additional dipole via multiple scattering from the $N$ initial dipoles. As with $\gamma_1$, a factor of $M_{ij}^{-1}\alpha_jG_{j,N+1}$ propagates the scattered field from the additional scatterer to the $i$th scatterer via all possible scattering paths not including the additional scatterer. The factor of $\alpha_{N+1}G_{N+1,i}$ then propagates the field back to the additional scatterer, from which it is scattered again, completing the loop. Summing over the number of loops yields a geometric series in terms of the single loop factor, and hence $\gamma_2$ can be expressed as a matrix inverse. This loop contribution is a self-interaction effect analogous to the surface dressing of polarizability. Finally, $\gamma_3$ accounts for the effect of scattering of the incident field onto the additional scatterer and therefore $\gamma_3$ describes the hotspot effect \cite{Cang2011ProbingImaging,Alonso-Gonzalez2012ResolvingSpots}. The incident field at the $j$th scatterer is multiply scattered to the $i$th scatterer, described by  $M_{ij}^{-1}$, and then propagated to the additional scatterer at $\bm{r}_{N+1}$, as is described by the final factor of $\alpha_{N+1}G_{N+1,i}$.

	{Optical sensing often aims to detect particles at a surface where functionalisation of the surface can allow for specificity. For this reason, we shall henceforth consider scattering configurations in which the $N$ initial scatterers are distributed over a planar surface at positions $z_i >0$. Physically, the scatterers could hence correspond to surface roughness features, bound receptors or nanoparticles, or nanostructures fabricated on a substrate. We also assume that the observation point $\bm{r}=(\bm{\rho},z)$ is taken in the far field as is the case in most sensing setups, which allows the form of $\gamma_1(\bm{r})$ to be greatly simplified.}
	Specifically, the far field Green's tensor $G_{\infty}$ is proportional to the 2D Fourier transform of the near field Green's tensor $\widetilde{G}(\bm{q},z,\bm{r}')$ with respect to the transverse position $\bm{\rho}=(x,y)$, i.e. \cite{Novotny2012PrinciplesNano-optics}
	\begin{equation}
		G_{\infty}(\bm{r},\bm{r}_i)=-ik\cos\theta\widetilde{G}(\bm{k}_\parallel,0;\bm{r}_i)\frac{e^{ikr}}{r} \label{eq:Ginf}
	\end{equation}
	where $k=n_bk_0$ is the wavenumber in the far field, $n_b$ is the refractive index at $z$, $\bm{k}_\parallel=(k_x,k_y)=k\sin\theta(\cos\phi,\sin\phi)$ is the 2D transverse component of the wavevector in the direction of observation and $(r,\theta,\phi)$ are the standard spherical coordinates of $\bm{r}$. Under the assumption of translational invariance in the transverse $(x,y)$ plane and that the scatterers all lie in the same bulk medium of dielectric constant $\varepsilon_d$, the far field Green's tensors for two different source positions are related. 
	{If the observation point is on the same side ($z>0$) as the scatterers then 
	\begin{align}\label{eq:G_ff_ref}
		G_\infty(\bm{r},\bm{r}_i)&=\left(G^{\text{dir}}_\infty(\bm{r},\bm{0})+G^{\text{ref}}_\infty(\bm{r},\bm{0})e^{2ik_zz_i}\right)\nonumber\\
		&\quad\quad\quad\quad\quad\quad\quad\quad\quad\times e^{-i(\bm{k}_\parallel\cdot\bm{\rho}_i+k_zz_i)}
	\end{align}
	whereas for observation points in the lower half space, for example if a thin film substrate is used, then 
	\begin{align}\label{eq:G_ff_trans}
	    G_\infty(\bm{r},\bm{r}_i)&=G^{\text{tr}}_\infty(\bm{r},\bm{0})e^{-i(\bm{k}_\parallel\cdot\bm{\rho}_j+k_zz_i)},
 	\end{align} 
	 where $k_z=\pm(\varepsilon_dk_0^2-k_\parallel^2)^{1/2}$, the upper (lower) sign is for observation points above (below) the interface and $G^{\text{dir}}_\infty$, $G^{\text{ref}}_\infty$ and $G^{\text{tr}}_\infty$ are the direct, reflected and transmitted components of the Green's tensor respectively \cite{Novotny2012PrinciplesNano-optics}}.  Under these assumptions, the Green's tensors in Eq.~\eqref{eq:gamma_1} cancel with the inverse Green's tensor factor, resulting in
	\begin{align}
		\gamma_1(\bm{k}_\parallel)&=I+\frac{k_0^2}{\varepsilon_0}\sum_{i,j=1}^N \Big[{R^\pm(z_i,z_{N+1})}e^{-i\bm{k}_\parallel\cdot(\bm{\rho}_i-\bm{\rho}_{N+1})} \nonumber\\
		&\quad\quad\quad\times e^{-ik_z(z_i-z_{N+1})}{M}_{ij}^{-1} \alpha_j {G}_{j,N+1}\Big].\label{eq:gamma_1_ff}
	\end{align}
	The function $R^\pm(z_i,z_{N+1})$ is derived in the appendix. Importantly, $R^-(z_i,z_{N+1})=I$ for observation points $z<0$ and $R^\pm(z_i,z_i)=I$ regardless of observation point. The remaining two enhancement factors do not depend on the observation point and thus do not differ between the far-field or near-field. 
	
	We have so far only considered the perturbation to the electric field, however most experimental setups measure the intensity of light, $\mathcal{I}=\abs{\bm{E}}^2$. The intensity perturbation $\delta \mathcal{I}=\abs{\bm{E}'}^2-\abs{\bm{E}}^2$ is therefore the typical signal in scattering based optical sensing, and can be related to the field perturbation through
	\begin{equation}
		\delta \mathcal{I}(\bm{r})=\abs{\delta\bm{E}(\bm{r})}^2+2\Re\left[\delta\bm{E}(\bm{r})\cdot\bm{E}^*(\bm{r})\right],
	\end{equation}
	where, in addition to the intensity of the perturbation $\delta\bm{E}$, there is a term corresponding to the interference between the field perturbation and the initial field. If, as will typically be the case for a large number of scatterers, $\abs{\bm{E}}\gg\abs{\delta\bm{E}}$, the interference term dominates and the intensity perturbation can be significantly larger than the dark field case where only the intensity scattered by the analyte particle is present. This principle is central to iSCAT and related techniques \cite{Taylor2019InterferometricScattering,Zhang2020PlasmonicKinetics,Berk2020TrackingSpeckle}, but it is not a multiple scattering effect and can be achieved equally well within a single scattering regime (and indeed typically is in iSCAT experiments), whether the interference is with other scattered fields or an external reference field. As such, this form of signal enhancement is independent of the scattering regime and different to the enhancement mechanisms we are considering. The phase difference between $\bm{E}$ and $\delta\bm{E}$ is random in both the single and multiple scattering regimes, so that the phase statistics of the interference term are essentially identical. The primary difference in the statistics of the interference term between single and multiple scattering lies in the different amplitudes $\abs{\delta\bm{E}}$.
	
	\subsection{Mean Enhancement Factors}\label{sec:theory3}
	For any given scattering configuration the value of each enhancement factor can be determined, however, it is valuable to characterise the distribution and average properties of the enhancement factors over the ensemble of different random configurations. In the following, the transverse positions $\bm{\rho}_i$ of the initial scatterers are assumed to be independently randomly distributed with uniform probability across a 2D planar region of area $L^2$ on the surface of a substrate, with the same height $z_i=z_s$ for $i\leq N$. Furthermore, the initial scatterers are assumed to be identical and to have the same orientation relative to the surface, whereby $\alpha_i=\alpha$ for $i\leq N$. Even for isotropic scatterers $\alpha$ may still be an anisotropic tensor due to surface dressing effects, however since the $N$ scatterers are at the same height, the surface dressing effect is identical for each scatterer. Note that the polarizability $\alpha_{N+1}$ is not restricted and may be different to the background scatterers. We do however limit attention to the case where $R^{\pm}(z_i,z_{N+1}) = I$ since this matches our simulations below and embodies all relevant physics in spite of the reduced mathematical complexity. Under these assumptions, $\langle\gamma_1\rangle$, $\langle S_2\rangle$ and $\langle\gamma_3\rangle$ can be calculated analytically, where $\gamma_2=(I-S_2)^{-1}$ and angled brackets denote averaging over realisations of the $N$ background scatterer positions $\bm{\rho}_i$. It should be noted, owing to the inverse relationship between $\gamma_2$ and $S_2$, their statistics have a more complicated relationship than the relationship between $\gamma_{1,3}$ and the corresponding sum terms appearing in Eqs.~\eqref{eq:gamma_1} and \eqref{eq:gamma_3}. Using Fourier analysis, $\gamma_{1,3}$ and $S_2$ can be expressed as
	\begin{widetext}
		\begin{align}
			\gamma_1(\bm{k}_\parallel)&=I+\frac{k_0^2}{\varepsilon_0}\int\frac{d^2 \bm{q}}{(2\pi)^2} {A}( \bm{k}_\parallel, \bm{q}) \alpha \widetilde{ {G}}( \bm{q};z_s,z_{N+1})e^{-ik_z( \bm{k}_\parallel)(z_s-z_{N+1})}\label{eq:gamma1_fourier}\\
			\gamma_3(\bm{E}_0)&=I+\frac{k_0^2}{\varepsilon_0}\int\frac{d^2 \bm{q}}{(2\pi)^2}\frac{d^2 \bm{q}'}{(2\pi)^2}\alpha_{N+1} \widetilde{ {G}}( \bm{q};z_{N+1},z_s)  {A}( \bm{q}, \bm{q}') \alpha  \widetilde{\bm{E}}_0( \bm{q}_2;z_s)e^{i \bm{q}'\cdot  \bm{\rho}_{N+1}}\frac{ \bm{p}_{0,N+1}^\dag}{ \abs{\bm{p}_{0,N+1}}^2} \label{eq:gamma3_fourier}\\
			{S}_2&=\left(\frac{k_0^2}{\varepsilon_0}\right)^2\int\frac{d^2 \bm{q}}{(2\pi)^2}\frac{d^2 \bm{q}'}{(2\pi)^2}\alpha_{N+1} \widetilde{ {G}}( \bm{q};z_{N+1},z_s)  {A}( \bm{q}, \bm{q}') \alpha \widetilde{ {G}}( \bm{q}';z_s,z_{N+1}) \label{eq:S2_fourier} ,
		\end{align}
	\end{widetext}
	where $\widetilde{f}(\bm{q})$ denotes the 2D Fourier transform of $f(\bm{\rho})$ such that $f(\bm{\rho})=\int\widetilde{f}(\bm{q})e^{i\bm{q}\cdot\bm{\rho}}d^2\bm{q}/(2\pi)^2$ and the function $A(\bm{q},\bm{q}')$ is defined by
	\begin{equation}\label{eq:A_def}
		{A}( \bm{q}, \bm{q}')=\sum_{i,j=1}^Ne^{-i \bm{q} \cdot( \bm{\rho}_i- \bm{\rho}_{N+1})} {M}_{ij}^{-1}e^{i \bm{q}'\cdot ( \bm{\rho}_j- \bm{\rho}_{N+1})}.
	\end{equation}
	In this form, the dependence of $\gamma_{1,3}$ and $S_2$ on the background scatterer positions is entirely described by $A$, such that their statistics are determined solely by the statistics of $A$. Accordingly, the means of Eqs. \eqref{eq:gamma1_fourier}--\eqref{eq:S2_fourier} can be calculated from $\langle A(\bm{q},\bm{q}')\rangle$. In order to calculate $\langle A\rangle$, we use the Neumann series  $(I-P)^{-1}=\sum_{l=0}^\infty P^k$ to expand $M_{ij}^{-1}$ as $M_{ij}^{-1}=I\delta_{ij}+\sum_{k=1}^\infty P_{ij}^k$ where
	\begin{align}\label{eq:k_order_Born}
		{P}^k_{ij}=\left(\frac{k_0^2}{\varepsilon_0}\right)^k\smashoperator[r]{\sum_{\substack{l_1, l_2,\ldots,l_{k-1}=1\\
					l_{i+1}\neq l_i\\
					l_1\neq i\\
					l_{k-1}\neq j}}^N}&\alpha  {G}_{il_1} \alpha  {G}_{l_1l_2} \alpha  {G}_{l_2l_3}
		\ldots\alpha  {G}_{l_{k-1}j}.
	\end{align}
	Physically, Eq.~\eqref{eq:k_order_Born} shows how $M_{ij}^{-1}$ corresponds to a sum over all scattering paths starting at the $j$th scatterer and ending at the $i$th scatterer, with $P^{k}_{ij}$ corresponding to the contribution from all paths visiting exactly $k$ scatterers. Each factor of $\alpha G_{l_il_{i+1}}$ propagates the field to the next scattering event. The $l_{i+1}\neq l_i$ exclusion arises  because a scattering path does not visit the same scatterer consecutively (as the self interaction is accounted for in $\alpha$). Inserting this expansion into $A(\bm{q},\bm{q}')$, the $p$th order contribution, denoted $A^{(p)}(\bm{q},\bm{q}')$ such that $A(\bm{q},\bm{q}')=\sum_{p=0}^\infty A^{(p)}(\bm{q},\bm{q}')$, is given by
	\begin{align}
		\!\!\!\!{A}^{(p)}(\bm{q},\bm{q}')&=\left(\frac{k_0^2}{\varepsilon_0}\right)^p\!\!\!\!\sum_{\substack{i,j,l_1, l_2,\\\ldots,l_{p-1}}}   
		\!\!\!
		e^{-i\bm{q}\cdot (\bm{\rho}_i-\bm{\rho}_{N+1})}\alpha   {G}_{il_1} \alpha   {G}_{l_1l_2} \alpha   {G}_{l_2l_3}\nonumber\\
		&\quad\,\ldots \alpha   {G}_{l_rl_{r+1}}\ldots \alpha    {G}_{l_{p-1}j}e^{i\bm{q}'\cdot (\bm{\rho}_j-\bm{\rho}_{N+1})}.
	\end{align}
	where henceforth the limits and exclusions from the sums will be left implicit. Replacing each Green's tensor with its Fourier decomposition allows the dependence on the scatterer positions to be included within an exponential factor as follows
	\begin{align}\label{eq:A_a}
			{A}^{(p)}(\bm{q},\bm{q}')=\int\prod_{b=1}^{p}&\frac{d^2\bm{q}_b}{(2\pi)^2}\frac{k_0^2}{\varepsilon_0}\alpha \widetilde{ {G}}(\bm{q}_b;z_s,z_s)\nonumber \\
			\times\!\!\!\!\!\sum_{i,j,l_1,\ldots,l_{p-1}}&\!\!\!\!e^{-i\bm{q}\cdot (\bm{\rho}_i-\bm{\rho}_{N+1})}e^{i\bm{q}_1\cdot (\bm{\rho}_i-\bm{\rho}_{l_1})}e^{i\bm{q}_2\cdot (\bm{\rho}_{l_1}-\bm{\rho}_{l_2})}\nonumber \\
			&\!\!\!\!\!\!\ldots e^{i\bm{q}_p\cdot (\bm{\rho}_{l_{p-1}}-\bm{\rho}_j)}e^{i\bm{q}'\cdot (\bm{\rho}_j-\bm{\rho}_{N+1})}.
	\end{align}
	The only random component of Eq.~\eqref{eq:A_a} is the exponential factors. Regrouping the exponents so that each $\bm{r}_{l_i}$ term is in one exponential factor allows the sum to be rewritten as
	\begin{align}\label{eq:exp_rearranged}
		\sum_{i,j,l_1,\ldots,l_{p-1}}&e^{i(\bm{q}_1-\bm{q}) \cdot\bm{\rho}_i}e^{i(\bm{q}_2-\bm{q}_1)\cdot \bm{\rho}_{l_1}}\ldots e^{i(\bm{q}_{c+1}-\bm{q}_{c})\cdot \bm{\rho}_{l_c}}\nonumber\\
		&\ldots e^{i(\bm{q}'-\bm{q}_p) \cdot\bm{\rho}_{j}}e^{i(\bm{q}-\bm{q}') \cdot\bm{\rho}_{N+1}}.
	\end{align}
	In general, there are terms in the sum in Eq.~\eqref{eq:exp_rearranged} where $l_i=l_j$ even when $i\neq j$, meaning each exponential factor is not necessarily independent of the others and hence cannot be averaged individually. Following a similar approach to that taken in Ref.~\cite{Martin-Mayor2001TheSystems}, we first consider only terms with no shared indices, where each $l_i$ is distinct. For these terms, each exponential $e^{i(\bm{q}_{c+1}-\bm{q}_{c})\cdot \bm{\rho}_{l_c}}$ can be averaged independently from the rest. Since there are $p+1$ different scatterers in such terms, there exist $N(N-1)(N-2)\ldots(N-p))\approx N^{p+1}$ terms in the sum with no repeated scatterers. Averaging a general function $f(\bm{\rho_i})$ over a scatterer position $\bm{\rho}_i$ corresponds to the integral $\langle f(\bm{\rho}_i)\rangle =\int f(\bm{\rho}_i)d^2\bm{\rho}_i/L^2$. Therefore, averaging over the $p+1$ different scatterer positions gives a factor of $(L^2)^{-(p+1)}$, so that the contribution of these distinct scatterer terms scales as $n^{p+1}$, where $n=N/L^2$ is the areal scatterer density. If we now consider the contribution of terms in the sum with 1 repeated scatterer (corresponding to scattering paths involving loops), meaning $p$ distinct scatterers are visited, choosing $p$ scatterers out of $N$ options gives $N(N-1)\ldots(N-(p-1))\sim N^{p}$ such terms. In this case, averaging over the $p$ scatterer positions give $(L^{2})^{-p}$, so that the contribution of these single repeated scatterer terms to the total sum is $\sim n^p$. It can be seen that the contribution of terms with $r$ repeated scatterers to the total sum in Eq.~\eqref{eq:exp_rearranged} scales as $n^{p+1-r}$. While methods to calculate the contribution from these loop paths exist \cite{Martin-Mayor2001TheSystems}, here we only take the leading order terms in $n$, i.e. the no loop contributions where all the indices $i,j,l_1,\ldots,l_{p-1}$ are distinct. Within this approximation,
	in the limit of large system size and scatterer number, $L\to\infty$ and $N\to\infty$, while keeping the scatterer density $n$ constant, the identity $\langle \sum_{j=1}^Ne^{i\bm{q}_\parallel\cdot \bm{\rho}_j}\rangle\to n(2\pi)^2\delta(\bm{q}_\parallel)$ can be applied for each summation index. After averaging, each exponential factor in Eq.~\eqref{eq:exp_rearranged} can therefore be replaced with a Dirac $\delta$-function. Thus, the $p$th order contribution to $A$ can be approximated by
	\begin{equation}
		\langle  {A}^{(p)}(\bm{q},\bm{q}')\rangle\approx n^{p+1}\left[\frac{k_0^2}{\varepsilon_0}\alpha  \widetilde{ {G}}(\bm{q};z_s,z_s)\right]^{p}(2\pi)^2\delta(\bm{q}-\bm{q}').
	\end{equation}
	Summing over $p$ hence gives
	\begin{equation}\label{eq:A_mean_full}
		\langle  {A}(\bm{q},\bm{q}')\rangle\approx n(2\pi)^2\delta(\bm{q}-\bm{q}')\left[ {I}-n\frac{k_0^2}{\varepsilon_0}\alpha \widetilde{ {G}}(\bm{q};z_s,z_s)\right]^{-1}.
	\end{equation}
	The means of Eqs. \eqref{eq:gamma1_fourier}--\eqref{eq:S2_fourier}, to leading order in $n$, therefore follow and are given by
	\begin{align}
		&\langle\gamma_1(\bm{k}_\parallel)\rangle=\left[I-n\frac{k_0^2}{\varepsilon_0}\alpha \widetilde{ {G}}(\bm{k}_\parallel;z_s,z_s)\right]^{-1}\label{eq:gamma1_avg}\\
		&\langle\gamma_3(\bm{E}_0)\rangle=I+n\frac{k_0^2}{\varepsilon_0}\int\frac{d^2\bm{q}}{(2\pi)^2}\alpha_{N+1} \widetilde{ {G}}(\bm{q};z_{N+1},z_s)\nonumber\\
		&\,\,\,\,\left[  I-n\frac{k_0^2}{\varepsilon_0}\alpha \widetilde{ {G}}(\bm{q},z_s,z_s)\right]^{-1}\!\!\!\alpha  \widetilde{\bm{E}}_0(\bm{q};z_s)e^{i\bm{q}\cdot \bm{\rho}_{N+1}}\frac{\bm{p}_{0,N+1}^\dag}{\abs{\bm{p}_{0,N+1}}^2}\label{eq:gamma3_avg}\\
		&\langle {S}_2\rangle=n\left(\frac{k_0^2}{\varepsilon_0}\right)^2\int\frac{d^2 \bm{q}}{(2\pi)^2}\alpha_{N+1} \widetilde{ {G}}( \bm{q};z_{N+1},z_s)\nonumber\\
		&\quad\quad\quad\left[ {I}-n\frac{k_0^2}{\varepsilon_0}\alpha \widetilde{ {G}}( \bm{q};z_s,z_s)\right]^{-1}\alpha \widetilde{ {G}}( \bm{q};z_s,z_{N+1}).\label{eq:S2_avg}
	\end{align}
	For the simple case of an incident (lossless) plane wave $\bm{E}_0=A_0\bm{\hat{\xi}}\exp(i\bm{k}^{\textrm{in}}\cdot\bm{r})$ and isotropic polarizabilities, $\langle\gamma_3\rangle$ reduces to a much simpler form, specifically
	\begin{equation}
		\langle\gamma_3(\bm{k}^{\textrm{in}}_\parallel)\rangle=\left[I-n\frac{k_0^2}{\varepsilon_0}\alpha \widetilde{ {G}}(\bm{k}^{\textrm{in}}_\parallel;z_s,z_s)\right]^{-1}\bm{\hat{\xi}}\bm{\hat{\xi}}^\dagger\label{eq:gamma3_avg_plane_wave}.
	\end{equation}
	The similar forms of $\langle\gamma_1\rangle$ and $\langle\gamma_3\rangle$ reflect the reciprocal symmetry present between scattering of an incoming plane wave scattering into an outgoing plane wave \cite{NiallSymmConstraints}.
	
	A notable feature of Eq.~\eqref{eq:gamma1_avg} is the divergence when $I-n(k_0^2/\varepsilon_0)\alpha \widetilde{ {G}}(\bm{k}_\parallel;z_s,z_s)$ is singular, or in the scalar case, when $n(k_0^2/\varepsilon_0)\alpha \widetilde{ G}(\bm{k}_\parallel;z_s,z_s)=1$. When this condition is close to being satisfied (i.e. $\det[I-n(k_0^2/\varepsilon_0)\alpha \widetilde{ {G}}(\bm{k}_\parallel;z_s,z_s)]$ is close to zero), the mean will become very large, suggesting the multiple scattering environment is significantly more sensitive to the addition of a scatterer than the single scattering environment. In the scalar case, $\abs{\langle\gamma_1\rangle}$ has a maximum value $\abs{\langle\gamma_1\rangle}_{\textrm{max}}>1$ provided $\Re[\alpha\widetilde{ {G}}(-\bm{k}_\parallel;z_s,z_s]>0$, occurring at a density $n_{\text{opt},1}$, where
	\begin{align}
		n_{\textrm{opt},1}&=\frac{\Re\left[\alpha\widetilde{G}(\bm{k}_\parallel;z_s,z_s)\right]}{\frac{k_0^2}{\varepsilon_0}\abs{\alpha\widetilde{G}(\bm{k}_\parallel;z_s,z_s)}^2}\label{eq:n_opt}\\
		\abs{\langle\gamma_1\rangle}_{\textrm{max}}&=\frac{\abs{\alpha\widetilde{G}(\bm{k}_\parallel;z_s,z_s)}}{\Im\left[\alpha\widetilde{G}(\bm{k}_\parallel;z_s,z_s)\right]}. \label{eq:max_enhance}
	\end{align}
	Analogous expressions for $n_{\textrm{opt},3}$ and $\abs{\langle\gamma_3\rangle}_{\textrm{max}}$ arise in the lossless case, replacing $\bm{k}_\parallel$ with $\bm{k}^{\textrm{in}}_\parallel$ in the argument of the Green's function. Physically, we can understand these conditions by considering the phase shifts involved in scattering. The plane wave component of the field scattered from one scatterer at wavevector $\bm{q}$ is phase shifted by $\arg[\alpha\widetilde{G}(\bm{q})]$ relative to the incident field. For any multiple scattering path, this phase shift is acquired at each scattering event, in addition to a propagation phase from travelling between scatterers. On averaging over realisations, the propagation phases cancel out, while the phase shift imparted by scattering events remains constant. When $\Im[\alpha\widetilde{G}(\bm{q})]=0$ and $\Re[\alpha\widetilde{G}(\bm{q})]>0$, there is no phase shift upon scattering and the averaged multiple scattering paths add up in phase, giving a maximum amplitude which, since the $N\to\infty$ limit has been taken, diverges as there are an infinite number of scattering paths in this case. In turn, a divergence of Eq.~\eqref{eq:max_enhance} results. Of course, any given realisation need not be close to the mean, and the random propagation phase can play a large role for any given realisation. As a result, it is important to study the statistics beyond simply the complex means, which we do numerically below.
	
	\section{Numerical Results}\label{sec:numerical}
	\subsection{Numerical Model}\label{sec:numerical1}
	In order to further study the statistical properties of the enhancement factors, Monte Carlo simulations were performed for scattering of SPPs propagating at a metal-dielectric interface (with dielectric constants $\varepsilon_m$ and $\varepsilon_d$ respectively) by nanoparticles in the dielectric near the surface (see inset of Fig.~\ref{fig:gold_dressed_mean_high_loss}). As discussed above, this choice of system is motivated by the use of SPP scattering in biological sensors \cite{Zhang2020PlasmonicKinetics,Berk2020TrackingSpeckle,Yang2018InterferometricExosomes}.
	Specifically, realisations of randomly distributed scatterers were generated and their corresponding scattered fields calculated by solving Eq.~\eqref{eq:Mp_p0} and using Eq.~\eqref{eq:coup_dip}. The simulation was repeated with an additional particle (cf.  Eq.~\eqref{eq:e2}) from which the field perturbation and individual enhancement factors were determined. 
	Notably, a scalar model can be used to describe SPP scattering \cite{Bozhevolnyi1998ElasticExperiment,Evlyukhin2005Point-dipoleLimitations}, with the scalar field corresponding to the out-of-plane component $E_z$ of the SPP field. When both $z$ and $z'$ are near the interface, the Green's function can be approximated as a cylindrical wave   \cite{Evlyukhin2005Point-dipoleLimitations,SoNdergaard,Sndergaard2003VectorialInteractions} given by
	\begin{equation}\label{eq:G_SPP}
		G_{\textrm{SPP}}(\bm{r},\bm{r}')=iA_0e^{-ak_\text{SPP}(z+z')}H_0^{(1)}(k_\text{SPP}\abs{\bm{\rho}-\bm{\rho}'})
	\end{equation}
	where  $a=(\varepsilon_d/(-\varepsilon_m))^{1/2}$, $A_0=ak_\text{SPP} /[2(1-a^4)(1-a^2)]$, $k_\text{SPP}$ is the complex SPP wavenumber with corresponding absorption length $l_\text{abs}=(2\Im[k_\text{SPP}])^{-1}$ and $H_0^{(1)}(x)$ is the zeroth order Hankel function  of  the first kind. Simulations were performed using this Green's function. The incident field was taken to be a decaying SPP plane wave of the form $E_{0,z}(x)=\Theta(x)\exp(ik_\text{SPP}x)$, where $\Theta(x)$ is the Heaviside step function {and we assume $z_{N+1} = z_s$}. Evaluating Eqs.~\eqref{eq:gamma1_avg}--\eqref{eq:S2_avg} with these assumptions  gives
	\begin{align}
		\langle\gamma_1(k_\parallel)\rangle&=\frac{k_\text{SPP}^2-k_\parallel^2}{k_\text{SPP}^2-k_\parallel^2+4n\mu}\label{eq:gamma1_avg_spp}\\
		\langle \gamma_3(x_{N+1})\rangle&=-\frac{2n\mu\exp\left[i(\widetilde{k}(n)-k_\text{SPP})x_{N+1}\right]}{(k_\text{SPP}-\widetilde{k}(n))\widetilde{k}(n)}\label{eq:gamma3_avg_spp}\\
		\langle S_2\rangle&=-\frac{\mu_{N+1}}{\pi}\log(1+\frac{4n\mu}{k_\text{SPP}^2})\label{eq:S2_avg_spp}
	\end{align}
	where we have defined $\widetilde{k}(n)=(k_\text{SPP}^2+4n\mu)^{1/2}$, $\mu=\alpha(k_0^2/\varepsilon_0) A_0\exp[{-2ak_\text{SPP}z_s}]$ and $\mu_{N+1}$ is defined analogously with $\alpha_{N+1}$ and $z_{N+1}$ replacing $\alpha$ and $z_s$. In addition, the SPP elastic scattering cross section $\sigma_\text{SPP}=4\abs{\mu}^2/\Re[k_\text{SPP}]$ and corresponding scattering mean free path $l_s=(n\sigma_\text{SPP})^{-1}$ can be defined for this model \cite{Bozhevolnyi1998ElasticExperiment,Evlyukhin2005Point-dipoleLimitations}. Note that the complex incident wavevector $k_\text{SPP}$ (i.e. the presence of absorption) means that $\langle \gamma_3\rangle$ does not take the form of Eq.~\eqref{eq:gamma3_avg_plane_wave}. 
	In order to study the role of absorption, simulations were performed at two different wavelengths. Firstly, the `low loss' case was simulated at $\lambda_0=650$~nm, for which $\varepsilon_d=1.77$ (corresponding to water) and $\varepsilon_m=-13.68+1.04i$ (corresponding to gold \cite{JohnsonRefractiveIndex}), meaning that $k_\text{SPP}=(1.42+0.008i)k_0$. The `high loss' case corresponded to $\lambda_0=600$nm, for which $\varepsilon_d=1.77$ (water) and $\varepsilon_m=-8.0+2.1i$ (gold) were taken whereby $k_\text{SPP}=(1.49+0.05i)k_0$. The absorption lengths were $9.9\lambda_0$ and $1.6\lambda_0$ respectively.
	In each case, the number of scatterers $N$ was fixed (700 for the `low loss' case and 800 for the `high loss' case), and they were randomly distributed in a square of sides $L$. To vary the scatterer density $n$, $L$ was varied between $L=9.3\lambda_0$ and $L=118\lambda_0$ in the low loss case and between $L=8\lambda_0$ and $L=30\lambda_0$ for the high loss case. Different sets of parameters were chosen for the two different wavelengths in order to ensure the density ranges in each case included both the single scattering and strong multiple scattering ($l_s<\lambda_0$) regimes. In all simulations performed, the additional scatterer was identical to the other scatterers ($\alpha_{N+1}=\alpha$) and added at the fixed position $\bm{r}_{N+1}=(0,0,z_s)$. All data points shown were calculated using 50,000 realisations of different scatterer positions unless otherwise stated.

	\subsection{Sensitivity Enhancements: Absorption Dependence} \label{sec:numerical2}
	\begin{figure*}[t]
		\centering
		\includegraphics[width=0.95\textwidth]{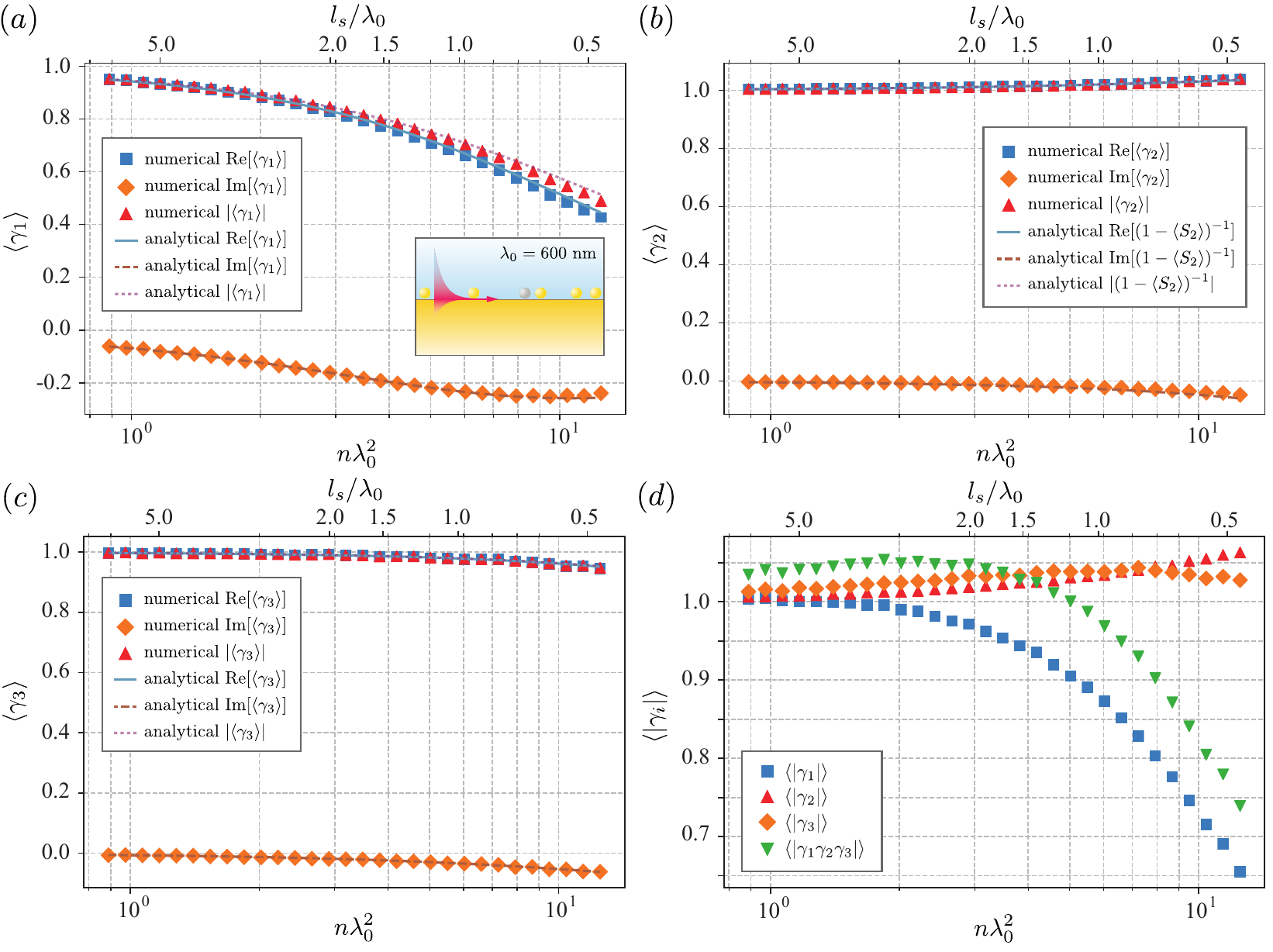}
		\caption{High loss, $\lambda_0=600$~nm, mean enhancements: dependence of the real part (blue $\square$), imaginary part (orange $\diamond$) and magnitude (red $\triangle$) of (a) $\langle\gamma_1\rangle$ (b) $\langle\gamma_2\rangle$ (c) $\langle\gamma_3\rangle$ on scatterer density $n$ (and mean free path $l_s$) for a 21.5~nm radius gold nanosphere sitting on the surface of gold interface (inset) as found from Monte-Carlo simulations. Corresponding analytic approximations are also shown (solid, dashed and dotted lines respectively). (d) Variation of the mean amplitudes of individual and total enhancement factors $\langle\abs{\gamma_1}\rangle$ (blue $\square$), $\langle\abs{\gamma_2}\rangle$ (red $\triangle$), $\langle\abs{\gamma_3}\rangle$ (orange $\diamond$) and $\langle\abs{\gamma_1\gamma_2\gamma_3}\rangle$ (green $\triangledown$).
		}
		\label{fig:gold_dressed_mean_high_loss}
	\end{figure*}
	Figs. \ref{fig:gold_dressed_mean_high_loss}(a)-(c) show the complex mean enhancements $\langle\gamma_i\rangle$  observed in the far field at $70^\circ$ to the surface normal in the backward direction ($\bm{k}_\parallel=-\varepsilon_d^{1/2}k_0\sin(70^\circ)\bm{\hat{x}}$) for $\lambda_0=600$~nm and assuming a polarizability $\alpha_{g1}$ corresponding to a 21.5nm radius gold sphere sitting on the gold surface.  The mean amplitudes $\langle\abs{\gamma_i}\rangle$ are also shown in Fig.~\ref{fig:gold_dressed_mean_high_loss}(d). The theoretical expressions (Eqs. \eqref{eq:gamma1_avg_spp} and \eqref{eq:gamma3_avg_spp}) are seen to describe $\langle\gamma_{1,3}\rangle$ well over the entire density range. Both $\langle\gamma_2\rangle$ and $\langle\gamma_3\rangle$ remain close to unity, as do the corresponding mean amplitudes, indicating that the effects of the associated multiple scattering paths are negligible. As a result, $\gamma_1$ is the dominant factor in the behaviour of the total mean amplitude enhancement $\langle\abs{\gamma_1\gamma_2\gamma_3}\rangle$ (Fig.~\ref{fig:gold_dressed_mean_high_loss}(d)), which scales very similarly to $\langle\abs{\gamma_1}\rangle$ and $\abs{\langle\gamma_1\rangle}$ (Fig.~\ref{fig:gold_dressed_mean_high_loss}(a)). 
	
\begin{figure*}[t]
		\centering
		\includegraphics[width=0.95\textwidth]{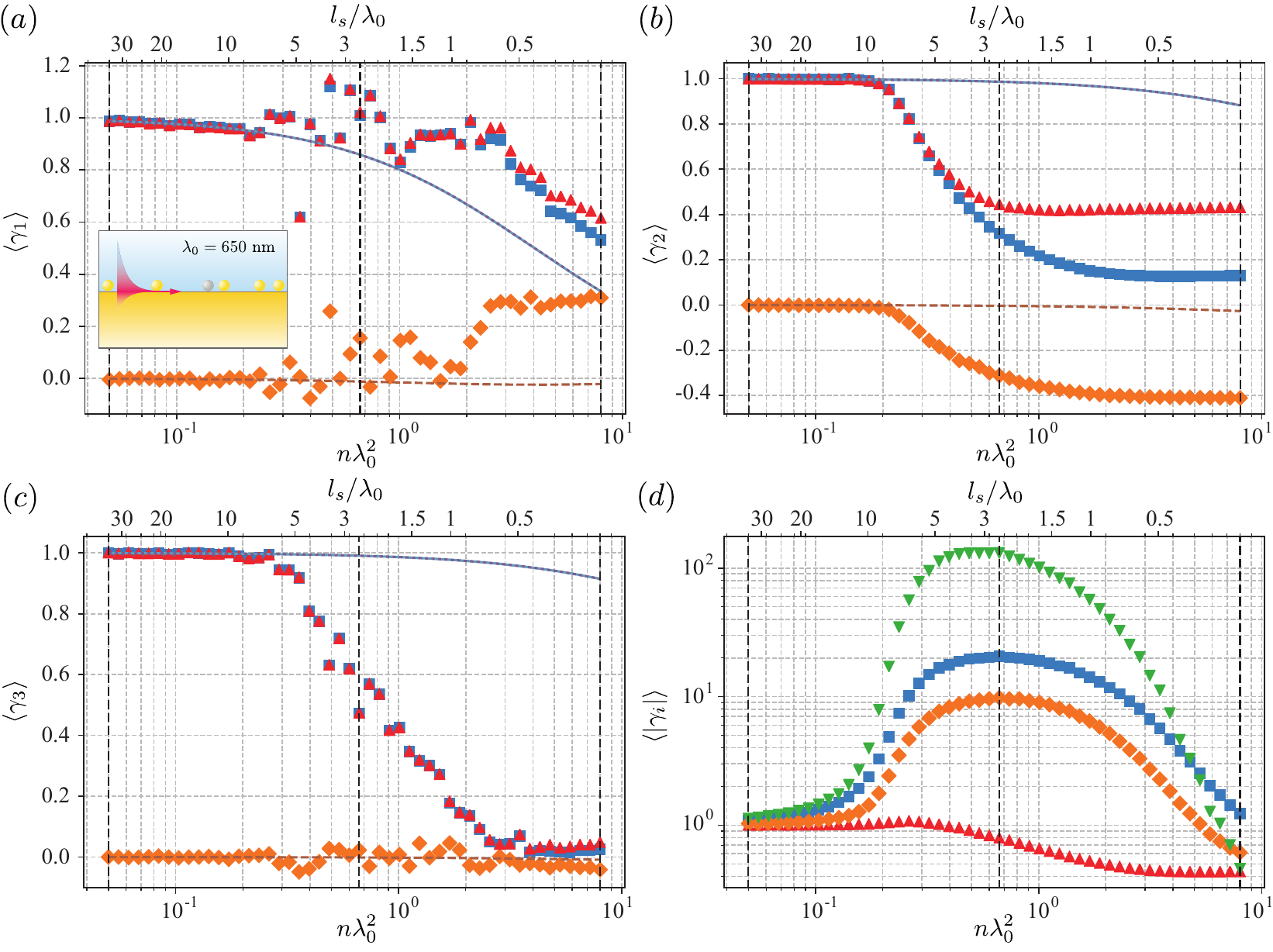}
		\caption{Low loss, $\lambda_0=650$~nm, mean enhancements: as Fig.~\ref{fig:gold_dressed_mean_high_loss} but for $\lambda_0=650$nm and polarizability $\alpha_{g2}$ corresponding to a $40$~nm gold sphere. Vertical dashed lines indicate densities at which probability distributions shown in Fig.~\ref{fig:gold_dressed_hist} are shown.}
		\label{fig:gold_dressed_mean_low_loss}
	\end{figure*}

\begin{figure*}[t]
		\centering
		\includegraphics[width=\textwidth]{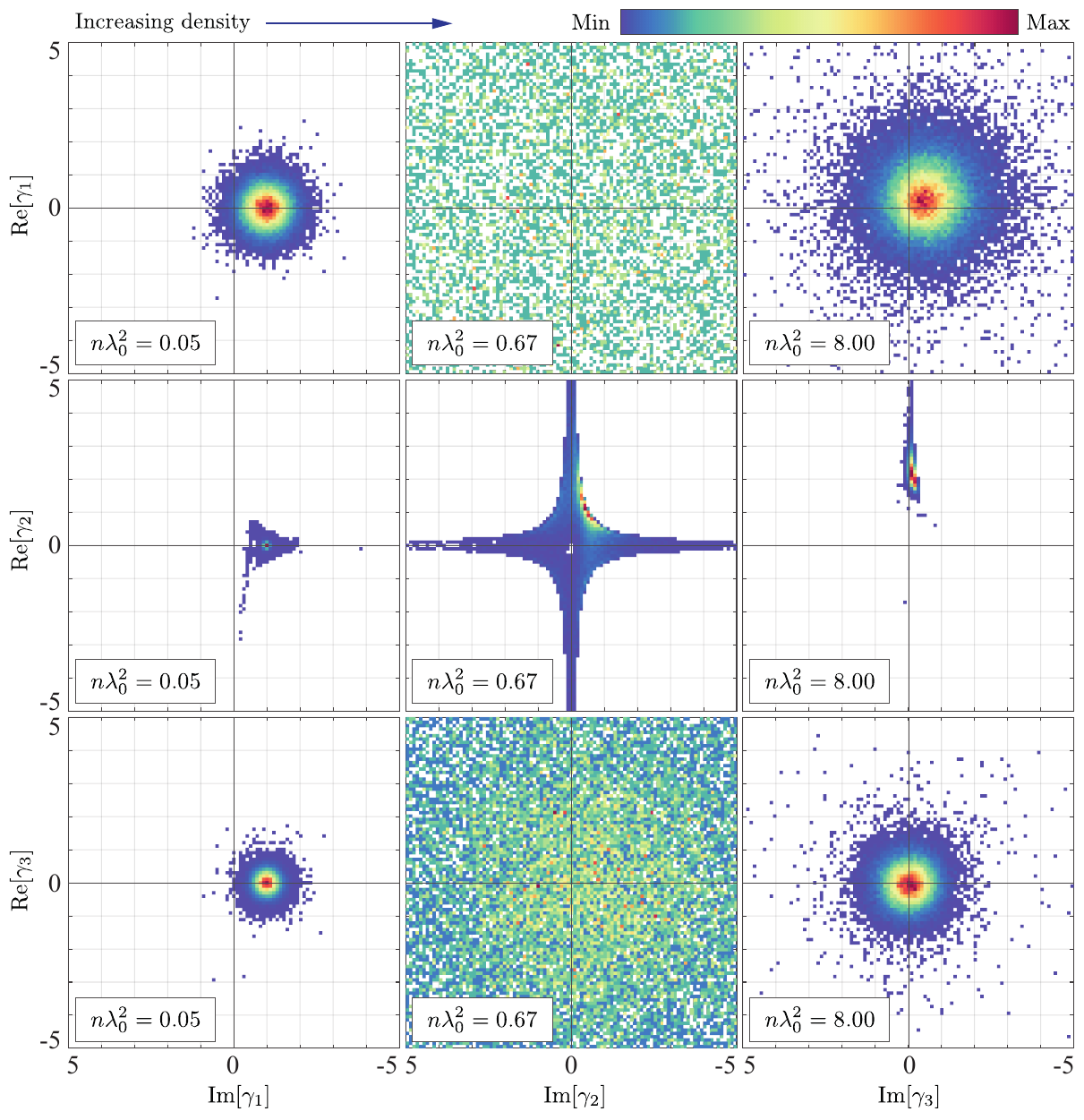}
		\caption{Histograms showing the relative frequency of $\gamma_1$ (top row), $\gamma_2$ (middle row) and $\gamma_3$ (bottom row) on the complex plane at densities $n\lambda_0^2=0.05$ (left column), $0.67$ (middle column) and $8.0$ (right column). Data shown corresponds to 50,000 realisations in the `low loss' case for polarizability $\alpha_{g2}$. Evolution of the distributions as scatterer density is increased can be seen in Supplementary Movies 1 and 2 \cite{SuppAnimations}.}
		\label{fig:gold_dressed_hist}
	\end{figure*}

	Equivalent plots for the low loss, $\lambda_0=650$nm, case with a polarizability $\alpha_{g2}$ equivalent to that of a $40$~nm gold sphere sitting on the surface and the same observation position are shown in Fig.~\ref{fig:gold_dressed_mean_low_loss}, from which a few significantly different features are evident. In the low loss case, the enhancement factors show greater deviation in the complex means from unity (Fig.~\ref{fig:gold_dressed_mean_low_loss}(a)--(c)), even at mean free paths of several wavelengths, which is unsurprising because the attenuation of propagating SPPs means the amplitude of multiple scattering paths are negligible when $l_s>l_\text{abs}$. The other significant difference between the low and high loss cases is in the mean of the absolute value of the enhancement factors (Fig.~\ref{fig:gold_dressed_mean_low_loss}(d)). The statistics of this quantity are explored in more detail in Ref. \cite{Berk2021}, but here we note that in the low loss case, $\langle\abs{\gamma_{1,3}}\rangle$ are very different from $\abs{\langle\gamma_{1,3}\rangle}$, by up to two orders of magnitude, whereas in the high loss case, the quantities are similar in value. Importantly, the low loss case allows for mean total amplitude enhancements $\langle\abs{\gamma_1\gamma_2\gamma_3}\rangle>1$, implying that multiple scattering increases the sensitivity, quite significantly, for a wide range of densities, whereas in the high loss case, multiple scattering only acts to decrease sensitivity on average. For the low loss case, the analytic results (Eqs.~\eqref{eq:gamma1_avg_spp}--\eqref{eq:S2_avg_spp}) still provide an accurate description at lower densities/longer mean free paths, however at higher densities, significant deviations are seen, particularly for $\langle\gamma_3\rangle$, indicating that the loop scattering paths ignored in the derivation of the average enhancements play a significant role. {Such loop paths are associated with weak localisation effects such as coherent back-scattering \cite{Akkermans2007MesoscopicPhotons}, which become significant at higher densities when $\Re[k_\text{SPP}]l_s\sim 1$.} 
	Furthermore, in the region where the mean amplitude grows large, the complex mean is slower to converge due to the larger variance in the underlying probability distribution (see Fig.~\ref{fig:gold_dressed_hist}) and hence larger statistical fluctuations are seen in the simulated data. Indeed, the results plotted between $n\lambda_0^2=0.21$ and $3.87$ in Fig. \ref{fig:gold_dressed_mean_low_loss} are averaged over 150,000 realisations in order to improve convergence. {Stronger Anderson localisation begins to play a role at the highest densities. The localisation length $\xi=l_s\exp\left(\pi \text{Re}[k_{\text{SPP}}]l_s/2\right)$ \cite{Sheng1995} becomes comparable to the system size for $l_s \approx 0.73\lambda_0$, at which point Anderson localisation means only scatterers within $\sim\xi$ couple strongly with each other. As a result, the effect of the added scatterer is reduced, explaining the decrease in mean amplitudes at the very highest densities.} 
	
	To study the underlying probability distributions in more detail we have plotted histograms of the relative frequency of the enhancement factors in the complex plane in Fig.~\ref{fig:gold_dressed_hist} for the low loss case at different densities. Supplementary Movies 1 and 2 show the complete density evolution of the distributions for both the high and low cases \cite{SuppAnimations}. In general, $\gamma_1$ and $\gamma_3$ appear to be distributed with rotational symmetry about their centres. Specifically, the standard deviations of the real and imaginary parts were found to typically be within 10\% of each other for both $\gamma_1$ and $\gamma_3$, although in some cases large outliers can cause significant differences. Similarly, the correlation coefficient between the phase and amplitude of the centred distribution $\gamma_{1,3}-\langle\gamma_{1,3}\rangle$ was never more than $\sim0.02$ across the density range considered. In contrast, $\gamma_2$, associated with loop scattering paths, has a more complicated locus on the complex plane, reminiscent of the previously studied eigenvalue distributions of Euclidean matrices arising in similar scattering studies \cite{Goetschy2013EuclideanPhysics}. In the low loss case the distributions of $\gamma_{1,3}$, while being narrow at low and high density become very broad for a range of intermediate densities. Thus, although the centre of the distribution remains close to the origin, the mean amplitudes $\langle\abs{\gamma_{1,3}}\rangle$ become very large as seen in Fig.~\ref{fig:gold_dressed_mean_low_loss}(d). In fact, the centre of the distributions, starting from $1$ at the lowest densities, move towards the origin with increasing density. This movement of the centre of the $\gamma_{1}$ distribution towards the origin is also seen in the high loss case (Fig.~\ref{fig:gold_dressed_mean_high_loss}(a)), however, the distribution remains tight around the centre over the full density range. Similarly, $\gamma_3$ retains the narrow width for the entire density range, although in this case the centre remains close to $1$. The similarity between the mean absolute values and the absolute value of the complex mean arises from these tight distributions. 
	
		\begin{figure*}[t]
		\centering
		\includegraphics[width=0.95\textwidth]{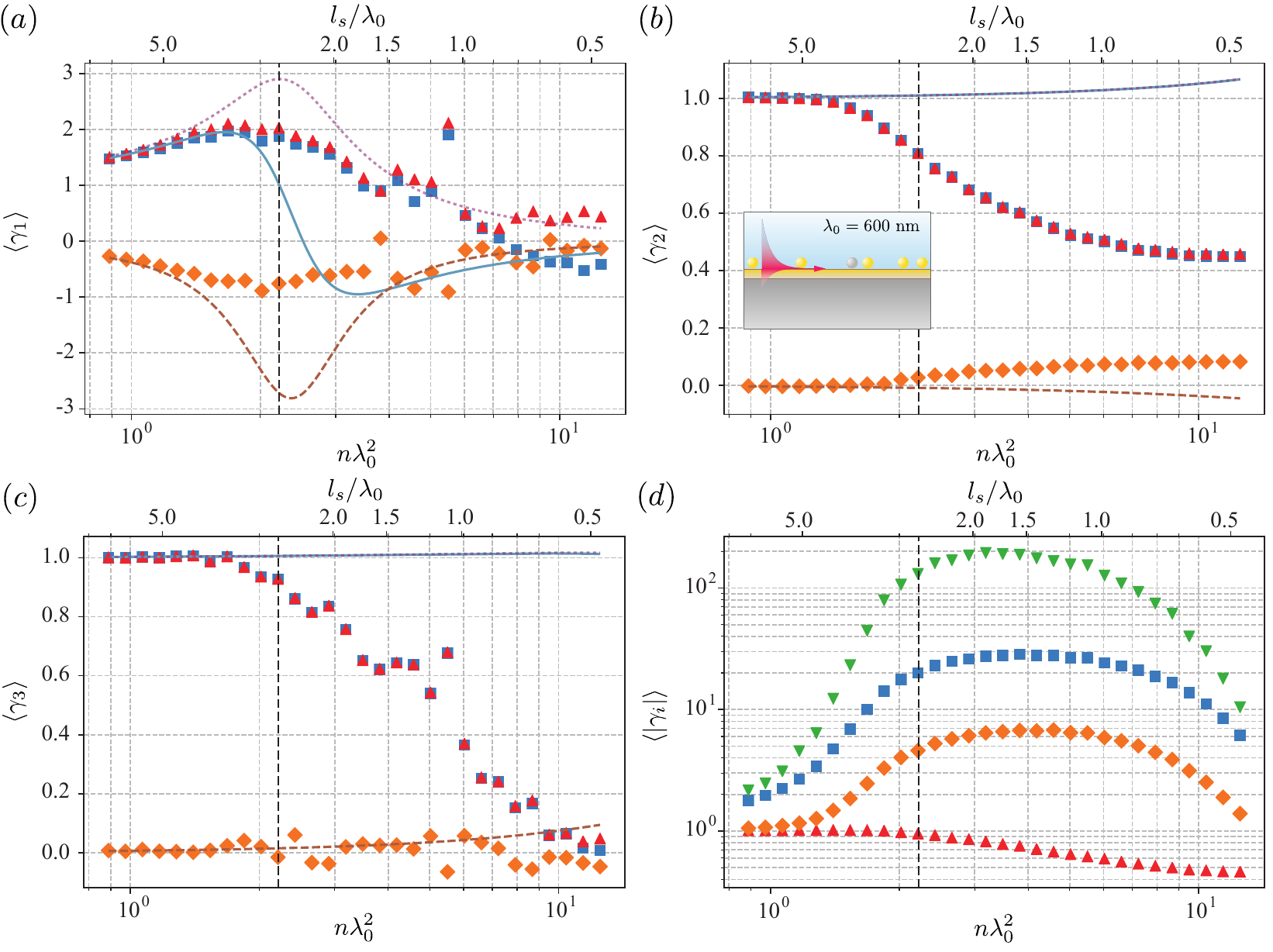}
		\caption{As Fig. \ref{fig:gold_dressed_mean_high_loss} but with an observation wavevector $\bm{k}_\parallel=-\Re[k_\text{SPP}]\bm{\hat{x}}$ and a phase shifted polarizability $\alpha=\alpha_{g1} e^{i\pi}$. Vertical dashed line indicates optimal scatterer density as predicted by Eq.~\eqref{eq:n_opt}.}
		\label{fig:near_opt_mean}
	\end{figure*}

	In order to understand the significant difference in the widths of the probability distributions for the high and low loss cases, we must consider the relative role of scattering and propagation phases along different multiple scattering trajectories. Each scattering path has an associated phase and amplitude which are determined by contributions from scattering events ($A_\text{scat}e^{i\Phi_\text{scat}}$) and from propagation between scattering events ($A_\text{prop}e^{i\Phi_\text{prop}}$), such that the enhancement factors are determined from the sum over all possible paths $\sim\sum_\text{paths}A_\text{scat}e^{i\Phi_\text{scat}}A_\text{prop}e^{i\Phi_\text{prop}}$. Changing realisations changes the propagation factors while the scattering contribution for a given sequence of scatterers is unchanged, since the scatterer positions change  but not their properties. When averaging over realisations, the random $\Phi_\text{prop}$ leads to cancellation of the propagation component and thus the complex mean simplifies to the sum of the deterministic  $A_\text{scat}e^{i\Phi_\text{scat}}$ factors arising from scattering events. Absorption means that scattering paths longer than $l_\text{abs}$ have a small amplitude $A_\text{scat}$ and hence contribute negligibly to the enhancement factors for that particular realisation. In the low loss case ($l_\text{abs}=9.9\lambda_0$), a large number of scattering paths several wavelengths long contribute. As the paths extend over multiple wavelengths, the phases $\Phi_\text{scat}$ are essentially uniform and random and thus the sum over scattering paths can give a significantly different result to the complex mean.  Conversely, in the high loss case, only a small number of scattering paths shorter than $l_\text{abs}=1.6\lambda_0$ contribute significantly to the enhancement factor. Furthermore, since the amplitude decay due to absorption occurs on the wavelength scale (the amplitude decays by $\sim20$\% over one SPP wavelength in the high loss case compared to $\sim2$\% in the low loss case), very short sub-wavelength scattering paths for which $\Phi_\text{prop}$ is close to zero will have significantly higher amplitude and contribute more to the total enhancement factors. As a result, the high loss case is close to the complex mean since the propagation has little effect. The behaviour of $\gamma_{1,3}$ in the high loss case is therefore dominated by the scattering phase shift.

	\subsection{Optimising Enhancements: Scatterer Dependence}\label{sec:numerical3}
		
	\begin{figure*}[t]
		\centering
		\includegraphics[width=\textwidth]{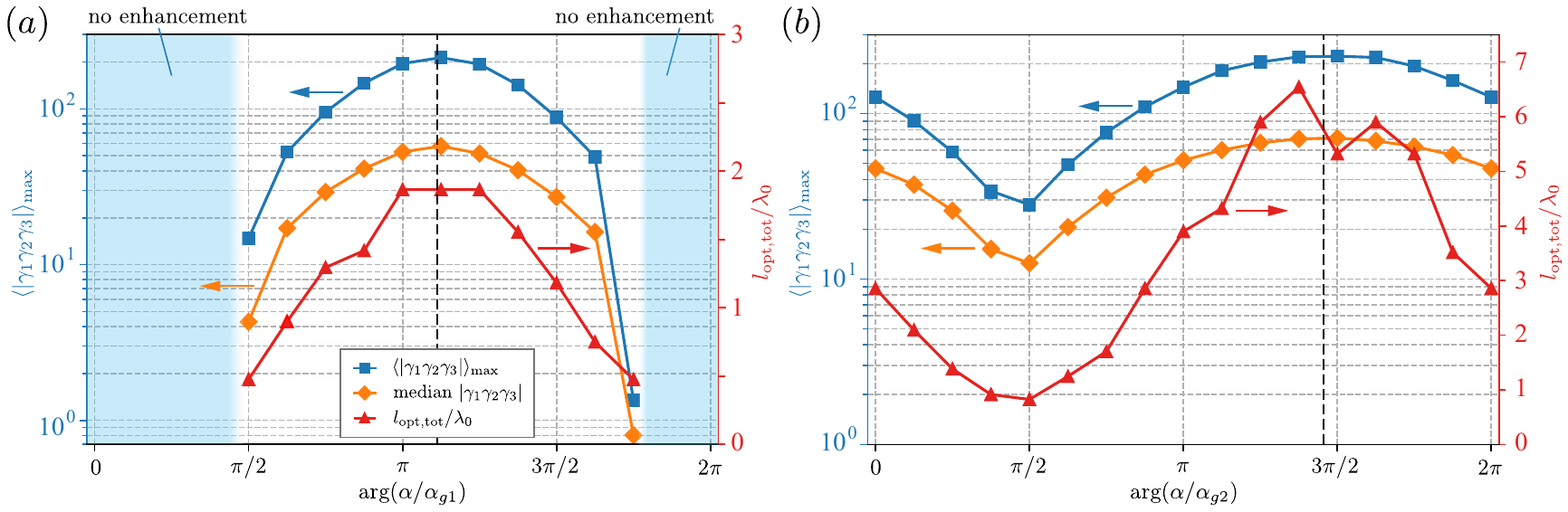}
		\caption{The maximum mean amplitude enhancement $\langle\abs{\gamma_1\gamma_2\gamma_3}\rangle$ (blue $\square$) and mean free path $l_{\text{opt,tot}}$ (red $\triangle$) at which it occurs for both high (a) and low (b) cases, as a function of the phase of $\alpha$ (or equivalently $\mu$) relative to that of a gold nanosphere on the surface. The observation point was taken in the leakage ring ($\bm{k}_\parallel=-\Re[k_\text{SPP}]\bm{\hat{x}})$. Median value of $\abs{\gamma_1\gamma_2\gamma_3}$ at $l_{\text{opt,tot}}$ is also shown (orange $\diamond$). Light blue shaded region indicates $\langle\abs{\gamma_1\gamma_2\gamma_3}\rangle\leq 1$, meaning that the single scattering case is optimum and multiple scattering always reduces sensitivity on average (these points are not plotted). The dashed black line denotes the optimum phase at which $\Im[\alpha\widetilde{G}]$ vanishes and Eq.~\eqref{eq:max_enhance} diverges.}
		\label{fig:alpha_phase_dep}
	\end{figure*}
	The conclusion that multiple scattering has a more pronounced effect in the low loss case in unsurprising, but the fact that the high loss case shows great sensitivity to the phase acquired in a scattering event, which is determined by the individual scatterer properties, is significant. To illustrate this, we consider a case close to the divergence condition of Eq.~\eqref{eq:max_enhance}. Since $\widetilde{G}(\bm{k}_\parallel)\propto1/(k_\text{SPP}^2-k_\parallel^2)$, the choice of observation point such that $\bm{k}_\parallel$ is close to the pole at $ k_\text{SPP}$, i.e. $k_\parallel=\Re[k_\text{SPP}]$, ensures a large $\abs*{\widetilde{G}}$ and thus the optimum density predicted by Eq.~\eqref{eq:n_opt} is reduced. In addition, it ensures the optimum density condition for $\gamma_1$ and the lossless version of $\gamma_3$ coincide (since the input and output wavevectors are equal). Such an observation point is not possible in a setup consisting of a single metal-dielectric interface since $\Re[k_\text{SPP}]>\varepsilon_d^{1/2}k_0$, however it is possible for SPPs excited in a thin gold film on a glass substrate  (see inset of Fig.~\ref{fig:near_opt_mean}). In particular, provided $n_\text{glass}k_0>\Re[k_\text{SPP}]$, an observation point taken in the leakage ring (a ring of directions in which light radiated to the far field is strongly confined \cite{Maier2007,Drezet2008LeakagePolaritons}) in the glass substrate  satisfies $k_\parallel=\Re[k_\text{SPP}]$. An observation position in the leakage ring furthermore has the additional benefit, from a sensing perspective, that the confinement of light  means detected signals are stronger. While such a thin film configuration alters the Green's function and surface dressing, the functional form of the SPP remains the same for points in the lower index dielectric near the surface of the gold film, with only the parameter values changed (i.e. $A_0$, $k_\text{SPP}$, $a$ and $\alpha$). We thus now consider such an observation position, keeping in mind that the parameters in the model will no longer correspond to the same physical properties.  Fig.~\ref{fig:near_opt_mean} shows the results from further simulations of the high loss case, analogous to those shown in Fig. \ref{fig:gold_dressed_mean_high_loss}, albeit assuming $\bm{k}_\parallel=-\Re[k_\text{SPP}]\bm{\hat{x}}$ and that the polarizability is phase shifted by $\pi$, i.e. $\alpha=\alpha_{g1} e^{i\pi}$. Note that since the amplitude of the polarizability is unchanged, the cross-section and mean free path are also unaltered. 
	The phase shift to $\alpha$ alters the absorption loss from a single scatterer, and also the phase difference between the scattered and incident field. The chosen phase means the divergence condition of Eq.~\eqref{eq:max_enhance} is nearly satisfied, i.e. the phase difference between the SPP incident on a scatterer and the SPPs radiated by the scatterer is small.
	We see significantly different behaviour in  Fig.~\ref{fig:near_opt_mean}  as compared to Fig.~\ref{fig:gold_dressed_mean_high_loss}. In particular, an optimum density $n_\text{opt}=3.18\lambda_0^{-2}$, at which $\langle\abs{\gamma_1\gamma_2\gamma_3}\rangle$ is maximised, is evident with a corresponding total amplitude enhancement of $\langle\abs{\gamma_1\gamma_2\gamma_3}\rangle=196$. The optimum density predicted from Eq.~\eqref{eq:n_opt} is $n_{\text{opt},1}=2.22\lambda_0^{-2}$. Critically, these large enhancements occur even with $l_s>l_\text{abs}$ when one might expect absorption to quench the effect of multiple scattering as was observed in Fig. \ref{fig:gold_dressed_mean_high_loss}. Results for the `low loss' case with $\alpha=\alpha_{g2} e^{3i\pi/4}$ (tuned near the divergence condition for the `low loss' parameters) were also obtained (not shown), however, in contrast to the high loss case, the behaviour of the means shows very little difference qualitatively from the results of Fig.~\ref{fig:gold_dressed_mean_low_loss} and with similar levels of enhancement observed.

	For the case shown in Fig.~\ref{fig:near_opt_mean} and its low loss counterpart, the probability distributions over the complex plane behave analogously to the behaviour shown in Fig. \ref{fig:gold_dressed_hist} (see also corresponding Supplementary Movies 3 and 4), with $\gamma_{1,3}$ showing both the initial broadening as scatterer density increases before contracting at higher densities, along with migration of the distribution centre from $1$ at low density towards $0$ at high density. The distribution of $\gamma_2$ appears relatively unchanged by the different observation position and tuning the phase of $\alpha$, maintaining the extended tails along the real and imaginary axes, while $\gamma_{1,3}$ again exhibit the same rotationally symmetric form.	
	
	In order to investigate the extent to which the divergence condition predicted in Eq.~\eqref{eq:max_enhance} holds, the maximum mean total absolute enhancement $\langle\abs{\gamma_1\gamma_2\gamma_3}\rangle_{\text{max}}$ and density $n_{\text{opt,tot}}$ (and corresponding mean free path $l_{\text{opt,tot}}$) at which it occurs were calculated numerically as the phase of $\mu$ was varied ($\abs{\mu}$ was again held constant and we assumed $\bm{k}_\parallel=-\Re[k_\text{SPP}]\bm{\hat{x}}$). While $\arg(\mu)$ is not dynamically tunable in general, it can be modified by changing
	various properties of the scatterers, for example their composition or geometry, or tuning the wavelength
	through a localised plasmonic resonance. More complex engineered scatterer structures such as core-shell nanospheres or nanorods allow further degrees of freedom for tuning $\alpha$. In addition, the phase of $\mu$ can be altered via its dependence on $z_s$ and use of index-matched spacer layers. Fig.~\ref{fig:alpha_phase_dep} shows the dependence of $\langle\abs{\gamma_1\gamma_2\gamma_3}\rangle_{\text{max}}$ and $l_{\text{opt,tot}}$ on $\arg(\mu)$, for both the high loss and low loss case. In the low loss case, we see that $\langle\abs{\gamma_1\gamma_2\gamma_3}\rangle_\text{max}$ is always achievable regardless of $\arg(\mu)$, with the value varying slightly with $\arg(\mu)$, albeit remaining $\sim 10^2$ for a broad range of phases. The optimum phase predicted from Eq.~\eqref{eq:max_enhance} coincides with the region where the enhancement is largest, and is also achieved at larger mean free paths (i.e. lower densities). Conversely, the high loss case has a range of $\arg(\mu)$ for which no enhancement is possible on average, since absorption quenches any multiple scattering enhancements. Tuning of $\arg(\mu)$ does nevertheless allow a similar level of enhancement to the low loss case to be achieved, with the divergence condition introduced by Eqs. \eqref{eq:n_opt} and \eqref{eq:max_enhance} providing a good predictor of the optimum phase. For the low loss case, long range scattering paths play a significant role as is discussed further in Ref.~\cite{Berk2021}.
	
	\section{Conclusion}
	To conclude, we have presented a general formalism to describe multiple scattering based enhancements to the field perturbation caused by adding an analyte particle into a random distribution of background scatterers. The approach presented is general and applicable to any wave scattering scenario, both vector and scalar, through appropriate choice of Green's tensor, for example scattering of acoustic waves or electromagnetic waves in free space, waveguides or photonic crystals \cite{Pinfield2017MultipleInteractions,Skipetrov3DMultScatt,WaveguideMultScatt,PhotCrystalMultipleScatt}. Three enhancement factors were derived, each arising from a different class of multiple scattering paths and their statistics were studied in the context of scattering of planar SPP waves.  Through a series of Monte-Carlo simulations we demonstrated that absorption can play an important role in the statistics of the enhancement factors, as it can quench long distance scattering paths. Supporting analytic calculations for the complex means of the enhancement factors were found to agree well when loop contributions were negligible. Whilst absorptive quenching was often seen to lead to an absence of any multiple scattering enhancement for high loss systems, the small propagation phases of short distance scattering paths imbues the system with a greater sensitivity to the scattering phase shift, and hence the individual scatterers. Consequently, we demonstrated that, by tuning the  polarizability of the background scatterers, a mean total enhancement of up to two orders of magnitude can be achieved.  Analytic expressions, capable of predicting the optimum polarizability, were also derived. Low loss systems were shown to exhibit contrasting behaviour. Specifically, it was found to always be possible to achieve an enhancement through appropriate tuning of the density of scatterers, regardless of the individual scatterer properties. Our results therefore demonstrate that multiple scattering can significantly enhance single particle detection, even in the presence of high losses, whilst insights gained can aid design of random scattering based nanostructured sensors, potentially enabling detection of weakly scattering particles such as single proteins or virions.

	\begin{acknowledgments}
		This work was funded by the Engineering and Physical Sciences Research
		Council (EPSRC) (1992728) and the Royal Society (UF150335).\\
	\end{acknowledgments}

	\appendix
	\section*{Appendix}
In this section we outline the derivation of the function $R^\pm(z_i,z_{N+1})$ appearing in Eq.~\eqref{eq:gamma_1_ff} defined through the equation
\begin{align}
	&G_\infty(\bm{r},\bm{r}_{N+1})^{-1}G_\infty(\bm{r},\bm{r}_i) \nonumber \\
	&\quad= R^\pm(z_i,z_{N+1}) e^{-i\bm{k}_\parallel\cdot(\bm{\rho}_i-\bm{\rho}_{N+1})}e^{-i{k}_z\cdot(z_i-z_{N+1})}.
\end{align}
 We assume that the upper interface of a planar stratified medium (such as a thin film structure) is located at $z=0$, whilst the lowest interface lies at a position $z=-d$. Recall that all scatterers are assumed to lie in the upper half-space $z_i>0$ for $i = 1,2,\ldots N+1$. We first note that from the translational invariance of the Green's function in the transverse plane, i.e. $G_\infty(\bm{r},\bm{r}_i) = G_\infty(\bm{r},\bm{z}_i) \exp(-i\bm{k}_\parallel\cdot \bm{\rho}_i)$ it follows immediately that $R^\pm(z_i,z_i) = I$.  Considering the more general case of observations positions lying in the lower half space, i.e. for $z < -d$, it also follows trivially that $R^-(z_i,z_{N+1})=I$ since there is only a transmitted component of the Green's function whereby from Eq. \eqref{eq:G_ff_trans}
\begin{align} 
	&G_\infty(\bm{r},\bm{r}_{N+1})^{-1}G_\infty(\bm{r},\bm{r}_i)\nonumber\\
	&\quad=
	G^{\text{tr}}_\infty(\bm{r},\bm{0})^{-1}G^{\text{tr}}_\infty(\bm{r},\bm{0})
	e^{-i\bm{k}_\parallel\cdot(\bm{\rho}_i-\bm{\rho}_{N+1})} 
	e^{-i{k}_z\cdot(z_i-z_{N+1})} 
	\nonumber\\
	&\quad=e^{-i\bm{k}_\parallel\cdot(\bm{\rho}_i-\bm{\rho}_{N+1})}e^{-i{k}_z\cdot(z_i-z_{N+1})}.
	\end{align}
In the reflection case, the Fourier space Green's tensor, for observation points above the source point $z>z_i>0$ is \cite{Novotny2012PrinciplesNano-optics}
\begin{align}\label{eq:G_fourier_reflection}
    \widetilde{G}(\bm{k}_\parallel;\bm{r}_i)&= \frac{i}{2k_z}  \widetilde{H}(\bm{k}_\parallel,z_i) e^{-i(\bm{k}_\parallel\cdot\bm{\rho}_i+k_zz_i)}
\end{align}
where 
\begin{align}
&\widetilde{H}(\bm{k}_\parallel,z_i)\nonumber\\	&\quad=(1+r_s(k_\parallel)e^{2ik_zz_i})\Gamma_s(\bm{k}_\parallel)+\Gamma_p(\bm{k}_\parallel)D(k_\parallel,z_i),
\end{align}
$D(k_\parallel,z_i)$ is a diagonal matrix given by
\begin{align}
D(k_\parallel,z_i)= I - \begin{bmatrix}
	1&0 & 0\\
	0& 1 &0 \\
	0& 0&-1
\end{bmatrix}r_p(k_\parallel)e^{2ik_zz_i} 
\end{align}
and $\Gamma_{s,p}$ are matrices projecting the source onto $s$ and $p$ polarized vectors, and can be expressed
\begin{align}
    \Gamma_{s,p}(\bm{k}_\parallel)=\bm{\hat{e}}_{s,p}(\bm{k}_\parallel)\bm{\hat{e}}_{s,p}^\dag(\bm{k}_\parallel).
\end{align}
The unit vectors $\bm{\hat{e}}_{s,p}(\bm{k}_\parallel)$ are the $s$ and $p$ polarized unit vectors for a plane wave of wavevector $\bm{k}_\parallel+k_z\bm{\hat{z}}$, given by
\begin{align}
    \bm{\hat{e}}_{s}(\bm{k}_\parallel)&=(-k_y,k_x,0)^T/k_\parallel\\
    \bm{\hat{e}}_{p}(\bm{k}_\parallel)&=(-k_xk_z,-k_yk_z,k_\parallel^2)^T/(\sqrt{\epsilon_d}k_0k_\parallel).
\end{align}
From Eq.~\eqref{eq:Ginf} we note that $G_\infty(\bm{r},\bm{r}_{N+1})^{-1}G_\infty(\bm{r},\bm{r}_i) = \widetilde{G}(\bm{k}_\parallel,\bm{r}_{N+1})^{-1}\widetilde{G}(\bm{k}_\parallel,\bm{r}_i)$ whereby 
\begin{align}
	&R^+(z_i,z_{N+1}) = \widetilde{H}(\bm{k}_\parallel,z_{N+1})^{-1}\widetilde{H}(\bm{k}_\parallel,z_i) .
\end{align}


\begin{thebibliography}{10}
		\newcommand{\enquote}[1]{``#1''}
		
		\bibitem{Taylor2019InterferometricScattering}
		R.~W. Taylor and V.~Sandoghdar, \enquote{{Interferometric Scattering
				Microscopy: Seeing Single Nanoparticles and Molecules via Rayleigh
				Scattering},} {{Nano Lett.}} \textbf{19}, 4827--4835
		(2019).
		
		\bibitem{Stetefeld2016DynamicSciences}
		J.~Stetefeld, S.~A. Mckenna, and T.~R. Patel, \enquote{{Dynamic light
				scattering: a practical guide and applications in biomedical sciences},}
		{{Biophys. Rev.}} \textbf{8}, 409--427 (2016).
		
		\bibitem{DWS88}
		D.~J. Pine, D.~A. Weitz, P.~M. Chaikin, and E.~Herbolzheimer,
		\enquote{Diffusing wave spectroscopy,} {{Phys. Rev.
				Lett.}} \textbf{60}, 1134--1137 (1988).
		
		\bibitem{Ye2019}
		Z.~Ye, X.~Wang, and L.~Xiao, \enquote{{Single-Particle Tracking with
				Scattering-Based Optical Microscopy},} {{Anal. Chem.}}
		\textbf{91}, 15327--15334 (2019).
		
		\bibitem{Shao2014}
		L.~Shao, X.-F. Jiang, X.-c. Yu, B.-b. Li, W.~R. Clements, F.~Vollmer, W.~Wang,
		Y.-F. Xiao, and Q.~Gong, \enquote{{Detection of Single Nanoparticles and
				Lentiviruses Using Microcavity Resonance Broadening},}
		{{Adv. Mater.}} \textbf{26}, 991 (2013).
		
		\bibitem{Li2021}
		N.~Li, T.~D. Canady, Q.~Huang, X.~Wang, G.~A. Fried, and B.~T. Cunningham,
		\enquote{{Photonic resonator interferometric scattering microscopy},}
		{{Nat. Commun.}} \textbf{12}, 1744 (2021).
		
		\bibitem{Xue2020}
		L.~Xue, H.~Yamazaki, R.~Ren, M.~Wanunu, A.~P. Ivanov, and J.~B. Edel,
		\enquote{{Solid-state nanopore sensors},} {{Nat. Rev.
				Materials}} \textbf{5}, 931--951 (2020).
		
		\bibitem{Homola2003PresentBiosensors}
		J.~Homola, \enquote{{Present and future of surface plasmon resonance
				biosensors},} {{Anal. Bioanal. Chem.}} \textbf{377},
		528--539 (2003).
		
		\bibitem{Baldrich2008}
		E.~Baldrich, O.~Laczka, F.~J. {Del Campo}, and F.~X. Mu{\~{n}}oz,
		\enquote{{Gold immuno-functionalisation via self-assembled monolayers: Study
				of critical parameters and comparative performance for protein and bacteria
				detection},} {{J. Immuno. Meth.}} \textbf{336}, 203--212
		(2008).
		
		\bibitem{JeffreyN.Anker2008BiosensingNanosensors}
		{Jeffrey N. Anker}, {W. Paige Hall}, {Olga Lyandres}, {Nilam C. Shah}, and
		{Jing Zhao {\&} Richard P. Van Duyne}, \enquote{{Biosensing with plasmonic
				nanosensors},} {{Nat. Mater.}} \textbf{7}, 442–453
		(2008).
		
		\bibitem{Zhang2020PlasmonicKinetics}
		P.~Zhang, G.~Ma, W.~Dong, Z.~Wan, S.~Wang, and N.~Tao, \enquote{{Plasmonic
				scattering imaging of single proteins and binding kinetics},}
		{{Nat. Methods}} \textbf{17}, 1010--1017 (2020).
		
		\bibitem{Zijlstra2012OpticalNanorod}
		P.~Zijlstra, P.~M. Paulo, and M.~Orrit, \enquote{{Optical detection of single
				non-absorbing molecules using the surface plasmon resonance of a gold
				nanorod},} {{Nat. Nanotech.}} \textbf{7},
		379--382 (2012).
		
		\bibitem{Raschke2003}
		G.~Raschke, {S. Kowarik}, {T. Franzl}, {C. S{\"{o}}nnichsen}, T.~A. Klar,
		J.~Feldmann, A.~Nichtl, and K.~K{\"{u}}rzinger, \enquote{{Biomolecular
				Recognition Based on Single Gold Nanoparticle Light Scattering},}
		{{Nano. Lett.}} \textbf{3}, 935--938 (2003).
		
		\bibitem{Taylor2017}
		A.~B. Taylor and P.~Zijlstra, \enquote{{Single-Molecule Plasmon Sensing:
				Current Status and Future Prospects},} {{ACS Sensors}}
		\textbf{2}, 1103--1122 (2017).
		
		\bibitem{Xue2019}
		T.~Xue, W.~Liang, Y.~Li, Y.~Sun, Y.~Xiang, Y.~Zhang, Z.~Dai, Y.~Duo, L.~Wu,
		K.~Qi, B.~N. Shivananju, L.~Zhang, X.~Cui, H.~Zhang, and Q.~Bao,
		\enquote{{Ultrasensitive detection of miRNA with an antimonene-based surface
				plasmon resonance sensor},} {{Nat. Commun.}}
		\textbf{10}, 28 (2019).
		
		\bibitem{Wen2015}
		Q.~Wen, X.~Han, C.~Hu, and J.~Zhang, \enquote{{Non-spectroscopic surface
				plasmon sensor with a tunable sensitivity},} {{Appl.
				Phys. Lett.}} \textbf{106}, 31113 (2015).
		
		\bibitem{Feng2012a}
		J.~Feng, V.~S. Siu, A.~Roelke, V.~Mehta, S.~Y. Rhieu, G.~T.~R. Palmore, and
		D.~Pacifici, \enquote{{Nanoscale plasmonic interferometers for multispectral,
				high-throughput biochemical sensing},} {{Nano Lett.}}
		\textbf{12}, 602--609 (2012).
		
		\bibitem{Bian2013}
		T.~Bian, B.~Z. Dong, and Y.~Zhang, \enquote{{A Broadband Nanosensor based on
				Multi-Interference of Surface Plasmon Polaritons},}
		{{Plasmonics}} \textbf{8}, 741--744 (2013).
		
		\bibitem{Zeng2015}
		B.~Zeng, Y.~Gao, and F.~J. Bartoli, \enquote{{Differentiating surface and bulk
				interactions in nanoplasmonic interferometric sensor arrays},}
		{{Nanoscale}} \textbf{7}, 166--170 (2015).
		
		\bibitem{Yang2018InterferometricExosomes}
		Y.~Yang, G.~Shen, H.~Wang, H.~Li, T.~Zhang, N.~Tao, X.~Ding, and H.~Yu,
		\enquote{{Interferometric plasmonic imaging and detection of single
				exosomes},} {{Proc. Natl. Acad. Sci. USA}} \textbf{115},
		10275--10280 (2018).
		
		\bibitem{Enoch2004}
		S.~Enoch, R.~Quidant, and G.~Badenes, \enquote{{Optical sensing based on
				plasmon coupling in nanoparticle arrays},} {{Opt.
				Express}} \textbf{12}, 3427 (2004).
		
		\bibitem{DalNegro2012}
		L.~{Dal Negro} and S.~Boriskina, \enquote{{Deterministic aperiodic
				nanostructures for photonics and plasmonics applications},}
		{{Laser Photon. Rev.}} \textbf{6}, 178--218 (2012).
		
		\bibitem{Lee2010}
		S.~Y. Lee, J.~J. Amsden, S.~V. Boriskina, A.~Gopinath, A.~Mitropolous, D.~L.
		Kaplan, F.~G. Omenetto, and L.~{Dal Negro}, \enquote{{Spatial and spectral
				detection of protein monolayers with deterministic aperiodic arrays of metal
				nanoparticles},} {{Proc. Natl. Acad. Sci. USA}}
		\textbf{107}, 12086--12090 (2010).
		
		\bibitem{LeMoal2009}
		E.~{Le Moal}, S.~L{\'{e}}v{\^{e}}que-Fort, M.-C. Potier, and E.~Fort,
		\enquote{{Nanoroughened plasmonic films for enhanced biosensing detection.}}
		{{Nanotechnology}} \textbf{20}, 225502 (2009).
		
		\bibitem{Szunerits2008}
		S.~Szunerits, V.~G. Praig, M.~Manesse, and R.~Boukherroub, \enquote{{Gold
				island films on indium tin oxide for localized surface plasmon sensing},}
		{{Nanotechnology}} \textbf{19}, 195712 (2008).
		
		\bibitem{Berkovits1994}
		R.~Berkovits and S.~Feng, \enquote{{Correlations in coherent multiple
				scattering},} {{Phys. Rep.}} \textbf{238}, 135--172
		(1994).
		
		\bibitem{Maystre1994}
		D.~Maystre and M.~Saillard, \enquote{{Localization of light by randomly rough
				surfaces: concept of localization},} {{J. Opt. Soc. Am.
				A}} \textbf{11}, 680--690 (1994).
		
		\bibitem{Boguslawski2017}
		M.~Boguslawski, S.~Brake, D.~Leykam, A.~S. Desyatnikov, and C.~Denz,
		\enquote{Observation of transverse coherent backscattering in disordered
			photonic structures,} {{Sci. Rep.}} \textbf{7}, 10439
		(2017).
		
		\bibitem{Segev2013}
		M.~Segev, Y.~Silberberg, and D.~N. Christodoulides, \enquote{Anderson
			localization of light,} {{Nat. Photonics}} \textbf{7},
		197--204 (2013).
		
		\bibitem{Shapiro1999NewMedia}
		B.~Shapiro, \enquote{{New Type of Intensity Correlation in Random Media},}
		{{Phys. Rev. Lett.}} \textbf{83}, 4733--4735 (1999).
		
		\bibitem{Skipetrov2000NonuniversalScattering}
		S.~Skipetrov and R.~Maynard, \enquote{{Nonuniversal correlations in multiple
				scattering},} {{Phys. Rev. B}} \textbf{62}, 886--891
		(2000).
		
		\bibitem{VanBeijnum2012b}
		F.~van Beijnum, J.~Sirre, C.~R{\'{e}}tif, and M.~P. van Exter,
		\enquote{{Speckle correlation functions applied to surface plasmons},}
		{{Phys. Rev. B}} \textbf{85}, 035437 (2012).
		
		\bibitem{Arnold1996}
		M.~Arnold and A.~Otto, \enquote{{Notes on localization of
				surface-plasmon-polaritons},} {{Opt. Commun.}}
		\textbf{125}, 122--136 (1996).
		
		\bibitem{Bozhevolnyi1996bbb}
		S.~I. Bozhevolnyi, A.~V. Zayats, and B.~Vohnsen, \enquote{{Weak Localization of
				Surface Plasmon Polaritons: Direct Observation with Photon Scanning Tunneling
				Microscope},} in \emph{Optics at the Nanometer Scale,}  (Springer
		Netherlands, 1996), pp. 163--173.
		
		\bibitem{Caze2012}
		A.~Caz{\'{e}}, R.~Pierrat, and R.~Carminati, \enquote{{Radiative and
				non-radiative local density of states on disordered plasmonic films},}
		{{Photon. Nanostruct.}} \textbf{10}, 339--344 (2012).
		
		\bibitem{Carminati2015}
		R.~Carminati, A.~Caz{\'{e}}, D.~Cao, F.~Peragut, V.~Krachmalnicoff, R.~Pierrat,
		and Y.~{De Wilde}, \enquote{{Electromagnetic density of states in complex
				plasmonic systems},} {{Surf. Sci. Reps.}} \textbf{70},
		1--41 (2015).
		
		\bibitem{Foreman2019a}
		M.~R. Foreman, \enquote{{Field Correlations in Surface Plasmon Speckle},}
		{{Sci. Rep.}} \textbf{9}, 8359 (2019).
		
		\bibitem{Bozhevolnyi2007}
		S.~I. Bozhevolnyi, \enquote{{Localization Phenomena in Elastic Surface Plasmon
				Polariton Scattering},} in \emph{Optical Properties of Nanostructured Random
			Media,}  (Springer Berlin Heidelberg, 2007), pp. 331--359.
		
		\bibitem{Tran2020UtilizingSensing}
		V.~Tran, S.~K. Sahoo, D.~Wang, and C.~Dang, \enquote{{Utilizing multiple
				scattering effect for highly sensitive optical refractive index sensing},}
		{{Sens. Actuators A Phys.}} \textbf{301}, 111776 (2020).
		
		\bibitem{MaumitaChakrabarti2015}
		{Maumita Chakrabarti}, {Michael Linde Jakobsen}, and {Steen G. Hanson},
		\enquote{{Speckle-based spectrometers},} {{Opt. Lett.}}
		\textbf{40}, 3264--3267 (2015).
		
		\bibitem{Katz2014}
		O.~Katz, P.~Heidmann, M.~Fink, and S.~Gigan, \enquote{{Non-invasive single-shot
				imaging through scattering layers and around corners via speckle
				correlations},} {{Nat. Photonics}} \textbf{8}, 784--790
		(2014).
		
		\bibitem{Lee2016}
		K.~R. Lee and Y.~K. Park, \enquote{{Exploiting the speckle-correlation
				scattering matrix for a compact reference-free holographic image sensor},}
		{{Nat. Commun.}} \textbf{7}, 1--7 (2016).
		
		\bibitem{Nieuwenhuizen1993}
		T.~M. Nieuwenhuizen and M.~C.~W. van Rossum, \enquote{{Role of a single
				scatterer in a multiple scattering medium},} {{Phys.
				Lett. A}} \textbf{177}, 102--106 (1993).
		
		\bibitem{Berkovits1991}
		R.~Berkovits, \enquote{{Sensitivity of the multiple scattering speckle pattern
				to the motion of a single scatterer},} {{Phys. Rev. B}}
		\textbf{43}, 8638--8640 (1991).
		
		\bibitem{Berkovits1990}
		R.~Berkovits and S.~Feng, \enquote{{Theory of speckle pattern tomography in
				multiple scattering media},} {{Phys. Rev. Lett.}}
		\textbf{65}, 3120--3123 (1990).
		
		\bibitem{Lancaster1998}
		D.~Lancaster and T.~M. Nieuwenhuizen, \enquote{{Scattering from objects
				immersed in a diffusive medium},} {{Physica A}}
		\textbf{256}, 417--438 (1998).
		
		\bibitem{Vynck2014}
		K.~Vynck, R.~Pierrat, and R.~Carminati, \enquote{{Polarization and spatial
				coherence of electromagnetic waves in uncorrelated disordered media},}
		{{Phys. Rev. A}} \textbf{89}, 013842 (2014).
		
		\bibitem{Berk2021}
		J.~Berk and M.~R. Foreman, {\enquote{{Theory of Multiple Scattering Enhanced Single Particle Plasmonic Sensing},}
		{{arxiv:2105.02798}}}  (2021).
		
		\bibitem{denOuter:93}
		P.~N. den Outer, T.~M. Nieuwenhuizen, and A.~Lagendijk, \enquote{Location of
			objects in multiple-scattering media,} {{J. Opt. Soc.
				Am. A}} \textbf{10}, 1209--1218 (1993).
		
		\bibitem{MultScattHolographLocalization}
		W.~Tahir, U.~S. Kamilov, and L.~Tian, \enquote{{Holographic particle
				localization under multiple scattering},} {{Adv.
				Photon.}} \textbf{1}, 1 -- 12 (2019).
		
		\bibitem{Suski2020FastMethod}
		D.~Suski, J.~Winnik, and T.~Kozacki, \enquote{{Fast multiple-scattering
				holographic tomography based on the wave propagation method},}
		{{Appl. Opt.}} \textbf{59}, 1397--1403 (2020).
		
		\bibitem{Kamilov2015LearningTomography}
		U.~S. Kamilov, I.~N. Papadopoulos, M.~H. Shoreh, A.~Goy, C.~Vonesch, M.~Unser,
		and D.~Psaltis, \enquote{{Learning approach to optical tomography},}
		{{Optica}} \textbf{2}, 517--522 (2015).
		
		\bibitem{Sun2018EfficientLearning}
		Y.~Sun, Z.~Xia, and U.~S. Kamilov, \enquote{{Efficient and accurate inversion
				of multiple scattering with deep learning},} {{Opt.
				Express}} \textbf{26}, 14678--14688 (2018).
		
		\bibitem{Moon2019}
		G.~Moon, T.~Son, H.~Lee, and D.~Kim, \enquote{{Deep learning approach for
				enhanced detection of surface plasmon scattering},}
		{{Anal. Chem.}} \textbf{91}, 9538--9545 (2019).
		
		\bibitem{Nishijima:12}
		Y.~Nishijima, L.~Rosa, and S.~Juodkazis, \enquote{Surface plasmon resonances in
			periodic and random patterns of gold nano-disks for broadband light
			harvesting,} {{Opt. Express}} \textbf{20}, 11466--11477
		(2012).
		
		\bibitem{Kim2010}
		K.~Kim, J.-W. Choi, K.~Ma, R.~Lee, K.-H. Yoo, C.-O. Yun, and D.~Kim,
		\enquote{{Nanoisland-Based Random Activation of Fluorescence for Visualizing
				Endocytotic Internalization of Adenovirus},} {{Small}}
		\textbf{6}, 1293--1299 (2010).
		
		\bibitem{Perumal2014}
		J.~Perumal, K.~V. Kong, U.~S. Dinish, R.~M. Bakker, and M.~Olivo,
		\enquote{{Design and fabrication of random silver films as substrate for SERS
				based nano-stress sensing of proteins},} {{RSC
				Advances}} \textbf{4}, 12995--13000 (2014).
		
		\bibitem{Frolov2013}
		L.~Frolov, A.~Dix, Y.~Tor, A.~B. Tesler, Y.~Chaikin, A.~Vaskevich, and
		I.~Rubinstein, \enquote{{Direct observation of aminoglycoside-RNA binding by
				localized surface plasmon resonance spectroscopy},}
		{{Anal. Chem.}} \textbf{85}, 2200--2207 (2013).
		
		\bibitem{Khurgin2015HowMetamaterials}
		J.~B. Khurgin, \enquote{{How to deal with the loss in plasmonics and
				metamaterials},} {{Nat. Nanotechnol.}} \textbf{10}, 2--6
		(2015).
		
		\bibitem{Sangu1999EffectMedia}
		S.~Sangu, T.~Okamoto, J.~Uozumi, and T.~Asakura, \enquote{{Effect of absorption
				on surface speckles in random media},} {{Waves Random
				Media}} \textbf{9}, 27--36 (1999).
		
		\bibitem{Genack1993IntensityMedia}
		A.~Z. Genack and N.~Garcia, \enquote{{Intensity Statistics and Correlation in
				Absorbing Random Media},} {{Europhys. Lett.}}
		\textbf{21}, 753--758 (1993).
		
		\bibitem{Pnini1991}
		R.~Pnini and B.~Shapiro, \enquote{{Intensity correlation in absorbing random
				media},} {{Phys. Lett. A}} \textbf{157}, 265--269
		(1991).
		
		
		
		\bibitem{Novotny1997InterferencePlasmons}
		L.~Novotny, B.~Hecht, and D.~W. Pohl, \enquote{{Interference of locally excited
				surface plasmons},} {{J. Appl. Phys.}} \textbf{81},
		1798--1806 (1997).
		
		\bibitem{Chaumet2005EfficientMethod}
		P.~C. Chaumet, A.~Rahmani, A.~Sentenac, and G.~W. Bryant, \enquote{{Efficient
				computation of optical forces with the coupled dipole method},}
		{{Phys. Rev. E}} \textbf{72}, 046708 (2005).
		
		\bibitem{Sndergaard2003VectorialInteractions}
		T.~S{\o}ndergaard and S.~I. Bozhevolnyi, \enquote{{Vectorial model for multiple
				scattering by surface nanoparticles via surface polariton-to-polariton
				interactions},} {{Phys. Rev. B}} \textbf{67}, 165405
		(2003).
		
		\bibitem{LakhtakiaCDA}
		{A.~Lakhtakia, \enquote{{Macroscopic theory of the coupled dipole approximation method},}
			{{Opt. Commun.}} \textbf{79}, 1 (1990).}
		
		\bibitem{Goetschy2011Non-HermitianTheory}
		A.~Goetschy and S.~E. Skipetrov, \enquote{{Non-Hermitian Euclidean random
				matrix theory},} {{Phys. Rev. E}} \textbf{84}, 011150
		(2011).
		
		\bibitem{Goetschy2013EuclideanPhysics}
		A.~Goetschy and S.~E. Skipetrov, \enquote{Euclidean random matrices and their
			applications in physics,}  (2013).
		
		\bibitem{Martin-Mayor2001TheSystems}
		V.~Martin-Mayor, M.~M{\'{e}}zard, G.~Parisi, and P.~Verrocchio, \enquote{{The
				dynamical structure factor in topologically disordered systems},}
		{{J. Chem. Phys.}} \textbf{114}, 8068--8081
		(2001).
		
		\bibitem{Mezard1999SpectraMatrices}
		M.~M{\'{e}}zard, G.~Parisi, and A.~Zee, \enquote{{Spectra of Euclidean random
				matrices},} {{Nuc. Phys. B}}  (1999).
		
		\bibitem{Cang2011ProbingImaging}
		H.~Cang, A.~Labno, C.~Lu, X.~Yin, M.~Liu, C.~Gladden, Y.~Liu, and X.~Zhang,
		\enquote{{Probing the electromagnetic field of a 15-nanometre hotspot by
				single molecule imaging},} {{Nature}} \textbf{469},
		385--388 (2011).
		
		\bibitem{Alonso-Gonzalez2012ResolvingSpots}
		P.~Alonso-Gonz{\'{a}}lez, P.~Albella, M.~Schnell, J.~Chen, F.~Huth,
		A.~Garc{\'{i}}a-Etxarri, F.~Casanova, F.~Golmar, L.~Arzubiaga, L.~E. Hueso,
		J.~Aizpurua, and R.~Hillenbrand, \enquote{{Resolving the electromagnetic
				mechanism of surface-enhanced light scattering at single hot spots},}
		{{Nat. Commun.}} \textbf{3}, 684 (2012).
		
		\bibitem{Novotny2012PrinciplesNano-optics}
		L.~Novotny and B.~Hecht, \emph{{Principles of Nano-Optics}} (Cambridge
		University Press, Cambridge, 2012), 2nd ed.
		
		\bibitem{Berk2020TrackingSpeckle}
		J.~Berk, C.~Paterson, and M.~R. Foreman, \enquote{{Tracking Single Particles
				using Surface Plasmon Leakage Radiation Speckle},} {{J.
				Light. Technol.}}  (2020).
		
		\bibitem{NiallSymmConstraints}
		N.~Byrnes and M.~R. Foreman, \enquote{Symmetry constraints for vector
			scattering and transfer matrices containing evanescent components: Energy
			conservation, reciprocity, and time reversal,} {{Phys.
				Rev. Research}} \textbf{3}, 013129 (2021).
		
		\bibitem{Bozhevolnyi1998ElasticExperiment}
		S.~I. Bozhevolnyi and V.~Coello, \enquote{{Elastic scattering of surface
				plasmon polaritons: Modeling and experiment},} {{Phys.
				Rev. B}} \textbf{58}, 10899--10910 (1998).
		
		\bibitem{Evlyukhin2005Point-dipoleLimitations}
		A.~B. Evlyukhin and S.~I. Bozhevolnyi, \enquote{{Point-dipole approximation for
				surface plasmon polariton scattering: Implications and limitations},}
		{{Phys. Rev. B}} \textbf{71}, 134304 (2005).
		
		\bibitem{SoNdergaard}
		T.~S{\o}ndergaard and S.~I. Bozhevolnyi, \enquote{{Surface plasmon polariton
				scattering by a small particle placed near a metal surface: An analytical
				study},} {{Phys. Rev. B}} \textbf{69}, 045422 (2004).
		
		\bibitem{JohnsonRefractiveIndex}
		P.~B. Johnson and R.~W. Christy, \enquote{Optical constants of the noble
			metals,} {{Phys. Rev. B}} \textbf{6}, 4370--4379 (1972).
		 
		 \bibitem{Akkermans2007MesoscopicPhotons}
	    {E.~Akkermans and G.~Montambaux, \emph{{Mesoscopic Physics of Electrons and
			Photons}} (Cambridge University Press, 2007), 1st ed.}
		
		 \bibitem{Sheng1995} 
	    {P. Sheng, \emph{Introduction to Wave Scattering, Localization and Mesoscopic Phenomena} (Academic Press Inc., New York, 1995).}
	
	
		\bibitem{SuppAnimations} See Supplemental Material at [URL will be inserted by publisher] for movie files showing evolution of probability distributions with scatterer density.
		
		\bibitem{Maier2007}
		S.~A. Maier, \emph{{Plasmonics: Fundamentals and Applications}} (Springer, 2007)
		
		\bibitem{Drezet2008LeakagePolaritons}
		A.~Drezet, A.~Hohenau, D.~Koller, A.~Stepanov, H.~Ditlbacher, B.~Steinberger,
		F.~R. Aussenegg, A.~Leitner, and J.~R. Krenn, \enquote{{Leakage radiation
				microscopy of surface plasmon polaritons},} {{Mater.
				Sci. Eng. B}} \textbf{149}, 220--229 (2008).
		
		\bibitem{Pinfield2017MultipleInteractions}
		V.~J. Pinfield and D.~M. Forrester, \enquote{{Multiple scattering in random
				dispersions of spherical scatterers: Effects of shear-acoustic
				interactions},} {{J. Acoust. Soc. Am.}} \textbf{141}, 649 (2017).
		
		\bibitem{Skipetrov3DMultScatt}
		S.~E. Skipetrov and I.~M. Sokolov, \enquote{Absence of Anderson localization of
			light in a random ensemble of point scatterers,} {{Phys.
				Rev. Lett.}} \textbf{112}, 023905 (2014).
		
		\bibitem{WaveguideMultScatt}
		L.~Tsang, H.~Chen, C.-C. Huang, and V.~Jandhyala, \enquote{Modeling of multiple
			scattering among vias in planar waveguides using Foldy–Lax equations,}
		{{Microw. Opt. Technol. Lett.}} \textbf{31}, 201--208
		(2001).
		
		\bibitem{PhotCrystalMultipleScatt}
		S.~Mazoyer, J.~P. Hugonin, and P.~Lalanne, \enquote{Disorder-induced multiple
			scattering in photonic-crystal waveguides,} {{Phys. Rev.
				Lett.}} \textbf{103}, 063903 (2009).
				
				
	   
	
	\end{thebibliography}
\end{document}